\documentclass[apj]{emulateapj}

\slugcomment{\it Accepted by ApJ --- 29 Jan 2006} 
\shorttitle{Vert. Struct. of Outer Milky Way HI Disk}
\shortauthors{Levine, Blitz, \& Heiles}

\begin{document}

\title{The Vertical Structure of the Outer Milky Way HI Disk}
\author{E.S. Levine\altaffilmark{1}, Leo Blitz\altaffilmark{1}, and Carl Heiles\altaffilmark{1}
}
\affil{Department of Astronomy, University of California at Berkeley, Mail Code
3411, Berkeley, CA 94720 USA}
\email{elevine@astron.berkeley.edu}

\begin{abstract}
We examine the outer Galactic HI disk for deviations from the $b=0\degr$ plane by constructing maps of disk surface density, mean height, and thickness. We find that the Galactic warp is well described by a vertical offset plus two Fourier modes of frequency 1 and 2, all of which grow with Galactocentric radius. Adding the $m=2$ mode  accounts for the large asymmetry between the northern and southern warps. We use a Morlet wavelet transform to investigate the spatial and frequency localization of higher frequency modes; these modes are often referred to as ``scalloping.'' We find that the $m=$ 10 and 15 scalloping modes are well above the noise, but localized; this suggests that the scalloping does not pervade the whole disk, but only local regions.
\end{abstract}


\section{Introduction}
Although the topography of the gas disk of the Milky Way has previously been  mapped \citep[and many others]{W1957, HJK1982,BT1986}, its complete Fourier structure has never been quantitatively described, though \citet{BM1998} have approximated its three lowest frequency terms. 
  Which spatial oscillation frequencies are most important? 
  Is the scalloping a local or global effect? 
  This paper constitutes an in depth analysis of the shape of the outer HI disk, as well as the first quantitative analysis of the scalloping. 
 
  The large-scale warp in the gas disk of the Milky Way has been known since 1957 \citep{B1957,K1957,W1957,KHC1957}.  
The warp has a large amplitude, rising to a height greater than 4 kpc  at a Galactocentric radius of 25 kpc in the northern data. It is also asymmetric; in the south the gas falls about 1 kpc below the plane before rising back to it. By observing other galaxies, it is possible to develop a general understanding of how warps behave. 
 \citet{B1991} found that at minimum half of all galaxies are warped, and that galaxies with smaller dark matter halo core radii are less likely to be warped.
 \citet{B1990} claimed that a warp's line of nodes starts out straight, and at a transition begins to advance in the direction of rotation, with some exceptions.
Another survey has shown that warps are common in galaxies with HI disks that are extended compared to their optical components, and are often asymmetric \citep{GSK2002}. In the Milky Way, many components other than HI also participate in the warp. A partial list includes  dust \citep{FBD1994}, CO \citep{WBB1990}, solar neighborhood stars \citep{D1998}, and IRAS point sources \citep{DS1989}.

 Many efforts have been directed toward understanding the warp  on a theoretical basis. Bending modes have long been suspected as the mechanism creating and maintaining the warp. Early work studying the evolution of  bending mode oscillations was hampered by a lack of knowledge regarding galactic halos, but showed that the shape of the density falloff near the edge of the disk  plays an important role in the stability of bending modes \citep{HT1969,T1983}. 
 The distribution of matter in the halo controls the ability of the disk to sustain long-lived bending wave warps, so studying the properties of the warp will  tell us about the shape of the halo  \citep{B1978,S1984}. \citet{SC1988} argued that bending mode oscillations are plausible when the halo and the self-gravity of the disk are taken into account, but \citet{BJD1998} showed that in this situation the warp will wind up within a few dynamical times. Previous work has largely been concerned with the existence and behavior of $m=1$ warps only, though \citet{S1995} demonstrated the stability of $m=0$ modes in an axisymmetric halo.

Several other mechanisms have been suggested as possibilities for creating and maintaining a warp \citep{KG2000}. Gravitational interaction with satellites such as the Magellanic Clouds is a   promising candidate \citep{B1957,W1998,WB2005}, but there is a longstanding debate of whether tidal effects are strong enough to produce the observed effect \citep{K1957}. Indeed, even in galaxies with companions similar to the LMC, tidal amplification may not be strong enough to account for the size of the warp \citep{GKD2002}. 
Accretion of matter onto the halo is another plausible cause \citep{JB1999}, or matter can accrete directly onto onto the disk and torque the gas orbits \citep{LBB2002,S2004}.  Intergalactic magnetic fields can act on a slightly ionized gas disk to produce a warp \citep{BFS1990}, or the intergalactic medium can excite a warp through a wind \citep{KW1959}.

Throughout this paper, we will refer to shorter wavelength (5--25 kpc) oscillations in azimuth as scalloping. \citet{GKW1960} first noticed a ``waviness'' in the gas layer of the inner Galaxy;  observational evidence  for scalloping turns out to be prominent in the outer Galaxy \citep{HJK1982,KBH1982}. \citet{F1983} investigated the possibility that the Milky Way scalloping results from the Kelvin-Helmholz instability, and \citet{S1995} suggested that the scalloping should only be present in the outer parts of Galactic disks. \citet{SF1986} find azimuthal corrugations in HI and other components along the spiral arms after removing the warp. In an N-Body simulation, \citet{EE1997} found evidence for spiral corrugations due to gravitational interaction with a satellite galaxy.
In this paper we will only investigate oscillation in Galactocentric azimuth, and not in radius.
 
   In \S 2 we describe the method of transforming the Heliocentric data cube into Galactocentric coordinates, and present maps of the surface density, average height off the $b=0\degr$ plane, and vertical thickness of the outer Galaxy. While these maps are constructed using new data, they are not significantly different from previous work.  In \S 3 we perform  global and local analyses on these maps to better understand the warp and the scalloping.  

\section{Method}\label{sec:proc}

\subsection{Data Processing}
We use the 21 cm Leiden/Argentine/Bonn (LAB) data  \citep{LAB,HB1997,BALMPK2005,ABLMP2000} to conduct  a quantitative study of  the warp and  scalloping. The LAB survey is a combination of the LDS data set \citep{HB1997} with Southern sky observations from the IAR \citep{ABLMP2000}; however, much attention has been paid toward ensuring a uniform data set. The data are corrected for stray radiation. The combined survey maps the entire sky within $-450\le v_r\le 400$ km s$^{-1}$ with a resolution of 1.3 km s$^{-1}$; this velocity range includes all of the gas in the Galaxy in circular rotation.  We used the Hanning smoothed  data, which have a velocity resolution of 1.9 km s$^{-1}$,  and used only data with  $|b|\le30\degr$. The signal of the warp in the north can be weakly traced beyond this elevation limit \citep{B1985}, but the vast majority of the warp signal is included within our $b$ range.

The LAB survey contains a large number of angularly small emission features, especially at high latitudes, which are not contiguous with the gas in the disk, and may not even be in circular rotation.  Features like these are particularly troublesome at larger Galactic latitudes since they can contain enough gas to  contaminate our calculations, especially at large Galactocentric radii.  These objects are removed with a median filter so we can focus on the gas in the disk. Points with brightness temperature $T_b\le 0$ are temporarily filled with the value $0.01$ K for this filtering only. We then cycle through the LAB data cube and 
calculate the median of each point and its 12 nearest neighbors  in $\ell$ and $b$ at the same line-of-sight velocity $v_r$ (a two dimensional diamond shaped filter); call this median $T_m(\ell,b,v_r)$. Any point with $T_b>10~T_m$ is replaced with 10 $T_m$. Two examples of objects  caught by this filter are M31 and NGC 6822.

The survey grid in $\ell,b,$ and $v_r$ is not convenient for analyzing Galactic properties. Ideally, our measurements would be equally spaced in the Galactocentric cylindrical coordinates $R,\phi,$ and $z$; we must interpolate a Galactocentric grid from the LSR-centered data. The Galactic azimuth $\phi$ is defined such that it converges with galactic longitude $\ell$ at large $R$. To convert from observed $\ell,b,$ and $v_r$ to $R,\phi$, and $z$ we use the following mapping functions:
\begin{eqnarray}\label{eqn:transform}
\ell&=&\sin^{-1}\left[\frac{R}{r'}\sin\phi\right]\nonumber\\
b&=&\tan^{-1}\frac{z}{r'}\nonumber\\
v_r&=&\sin\ell\cos b\left[\frac{R_0}{R}\Theta(R)-\Theta_0\right]\nonumber\\
&&+v_\Pi(R)\cos\phi\left(1-\frac{R_0^2}{R^2}\sin^2\ell\right)^{1/2}\cos b.
\end{eqnarray}
Here, $\Theta(R)$ is the Galactic rotation curve, which we assume to be 220  km s$^{-1}$ everywhere \citep{BB1993}. $\Theta_0$ and $R_0$ are 220 km s$^{-1}$ and 8.5 kpc, respectively. $\bf{r}$ is the vector connecting the Sun's location to the point under investigation; $\bf{r'}$ is the projection of this vector onto the plane of the disk (with magnitude $r'$).

 Equation \ref{eqn:transform} is the transformation that results from an assumption of elliptical gas orbits with major axis
 along the $\phi=90\degr, 270\degr$ line. To derive this, we assume the gas moves on the orbit:
 \begin{eqnarray}\label{eqn:ellipse}
 v_\phi&=&220~\mathrm{km~s^{-1}}\nonumber\\
 v_R&=&v_\Pi(R)\cos\phi.
 \end{eqnarray}
$v_R$ is the magnitude of the velocity in the Galactocentric radial direction. $v_\Pi$ is a parameterization of the ellipticity of the orbit, which is free to vary with Galactocentric radius; we discuss our method of calculating $v_\Pi(R)$ in the Appendix. At all points, $v_R/v_\phi<0.1$. These equations are  simply the epicyclic approximation for an orbit with epicyclic frequency 1 and  the angle of ellipse orientation fixed.  Although the gas orbits in the Galaxy are not likely to correspond to the fixed ellipse orientation we describe, this configuration minimizes the correction to $v_r$ for gas far from $\ell=0\degr$ or $180\degr$ (see  \citet{V1999} for more detail on elliptical gas orbits). 
Without a correction for radial motion of the gas,  there is a large asymmetry between the surface densities at Galactic longitudes on either side of $\ell =0\degr$ and $\ell =180\degr$ \citep{HJK1982}. This must be taken into account, or features in these two regions such as the surface density will appear discontinuous  and distorted. Assuming an outward velocity for the Local Standard of Rest (LSR) will correct this discontinuity to some degree \citep{K1962,KW1965}. However, it seems that the best fit for the motion relative to the LSR  changes with radius, implying that the effect is global, rather than local \citep{BS1991}. To reduce the magnitude of the discontinuity using gas orbits, one can use a Galactocentric radial velocity roughly of the form $\cos\phi$ (or $\cos\ell$) \citep{KT1994}. 

We  exclude  all points that lie within $345\degr\le\ell\le15\degr$ or $165\degr\le\ell\le195\degr$. Points in these two wedges have velocities along the line of sight that are too small with respect to their random velocities to establish reliable distances. All points in this region are set to $T_b = 0$.

Using (\ref{eqn:transform}) we construct a Galactocentric grid $T_b(R,\phi,z)$  by trilinear-interpolating from the grid $T_b(\ell,b,v_r)$.  We do this by calculating the coordinates of an $(R,\phi,z)$ point in $(\ell,b,v_r)$ space, and interpolating from $T_b(\ell,b,v_r)$. The resolution of the Galactocentric grid is set by the spacing of the LSR centered grid, but in this paper we are not interested in  small-scale disk structure. A grid of 100 points in 10 kpc $\le R \le$ 30 kpc, 350 points in $-\pi\le\phi\le \pi$, and 141 points in -20 kpc$\le z \le$ 20 kpc gives us sufficient resolution to answer the questions we are interested in.

 Undersampled grids fail to utilize all of the information in the data; our choice of grid spacing is both is an undersampling and an oversampling of the information in $T_b(\ell,b,v_r)$ depending on the position in the disk. Consider  two points near $R=10$ kpc and $\ell = 15\degr$, where we have poor Galactocentric resolution in $\phi$. The LAB survey has $\Delta \ell = 0.5\degr$; at this location this corresponds to $\Delta\phi\approx0.9\degr$. The spacing in our Galactocentric grid is larger than $1\degr$, thus we are undersampled everywhere in the $\phi$ dimension. In \S \ref{sec:lomb} we will perform an azimuthal frequency analysis of each ring; none of the frequencies we examine approach the Nyquist frequency of the data.  Near the midplane at $R=30$ kpc and $\ell =15\degr$, lines of constant $b$  are separated by around 300 pc, and near the top of the grid they are separated by around 400 pc; there is some oversampling in the $z$ dimension by no more than a factor of 2. The most severe case of oversampling is in the $R$ dimension, where along $\ell = 15\degr$ we have only 14 velocity resolution elements in our $R$ range for the 100 grid points. On the other hand, at $\ell=90\degr$, there are more than 60 resolution elements in our range.  We have chosen a grid spacing in $R$ that oversamples to a varying degree depending on location in the disk.

From $T_b(R,\phi,z)$, we can recover $\rho(R,\phi,z)$ using the method outlined in \citet{K1968}. We assume $T_s$, the spin temperature, is 155 K everywhere. This is slightly higher than the maximum brightness temperature found in the LAB survey in the region we are concerned with. Choosing a larger number to force the optically thin limit ($T_b\ll T_s$) makes a difference only in a small number of areas in the inner radii of our grid. The vast majority of the points are optically thin with any reasonable $T_s$, and are not affected by this choice. 
The transformation to a density grid depends on $|dv_r/dr|$. Using elliptical orbits makes calculating this quantity slightly more difficult than with a flat rotation curve. Although the full derivative can be written analytically, we just calculate it numerically.
Points with $T_b<0$ are set to $\rho =0$.

\subsection{Surface Density and Mean Height maps}

The grid $\rho(R,\phi,z)$ contains information about the density of HI in the Galaxy, minus whatever has been removed by the median filter and the excluded regions. Previous studies have proceeded by calculating a mean height $\bar{z}(R,\phi)$ for the gas. However, the Galaxy is a complicated place that contains a variety of HI structures in addition to  the disk. In particular, there are many extended clouds located near the disk as well as spurs that split off from the disk. None of these will have been removed by the median filter, which acts only on comparatively small areas of the sky. \citet{V1999} developed an alternative method to remove some of these features; he masked out a map of high velocity cloud complexes. 
Since we are only interested in the shape of the disk itself, these additional  components must be filtered out before  our calculation of the mean height. We perform a dispersion filter that operates as follows.
\begin{enumerate}
\item Calculate the total mass surface density\begin{equation}M(R,\phi)=\sum_{i=1}^{N=141} \rho(R,\phi,z_i)\Delta z\end{equation}
where $\Delta z$ is the $z$ bin size.
\item Calculate \begin{equation}\bar{z}(R,\phi)=\frac{\sum_{i=1}^{N=141} z_i \rho(R,\phi,z_i) \Delta z}{M(R,\phi)}.\end{equation}
\item Calculate the second moment\begin{equation}d^2(R,\phi)=\frac{\sum_{i=1}^{N=141}[z_i-\bar{z}(R,\phi)]^2 \rho(R,\phi,z_i) \Delta z}{M(R,\phi)}.\end{equation}
\item Run through each point in $(R,\phi,z)$ space.
For any point that does not lie within $2d$ of $\bar{z}$, set $\rho = 0$. Call the grid that results from this dispersion filter $\rho_d(R,\phi,z)$.
\end{enumerate}
For step 4, we experimented with several different cutoffs (in the range $1-3d$); our results do not depend strongly on which cutoff we choose.

\begin{figure*}
\includegraphics[angle=90,scale=.88]{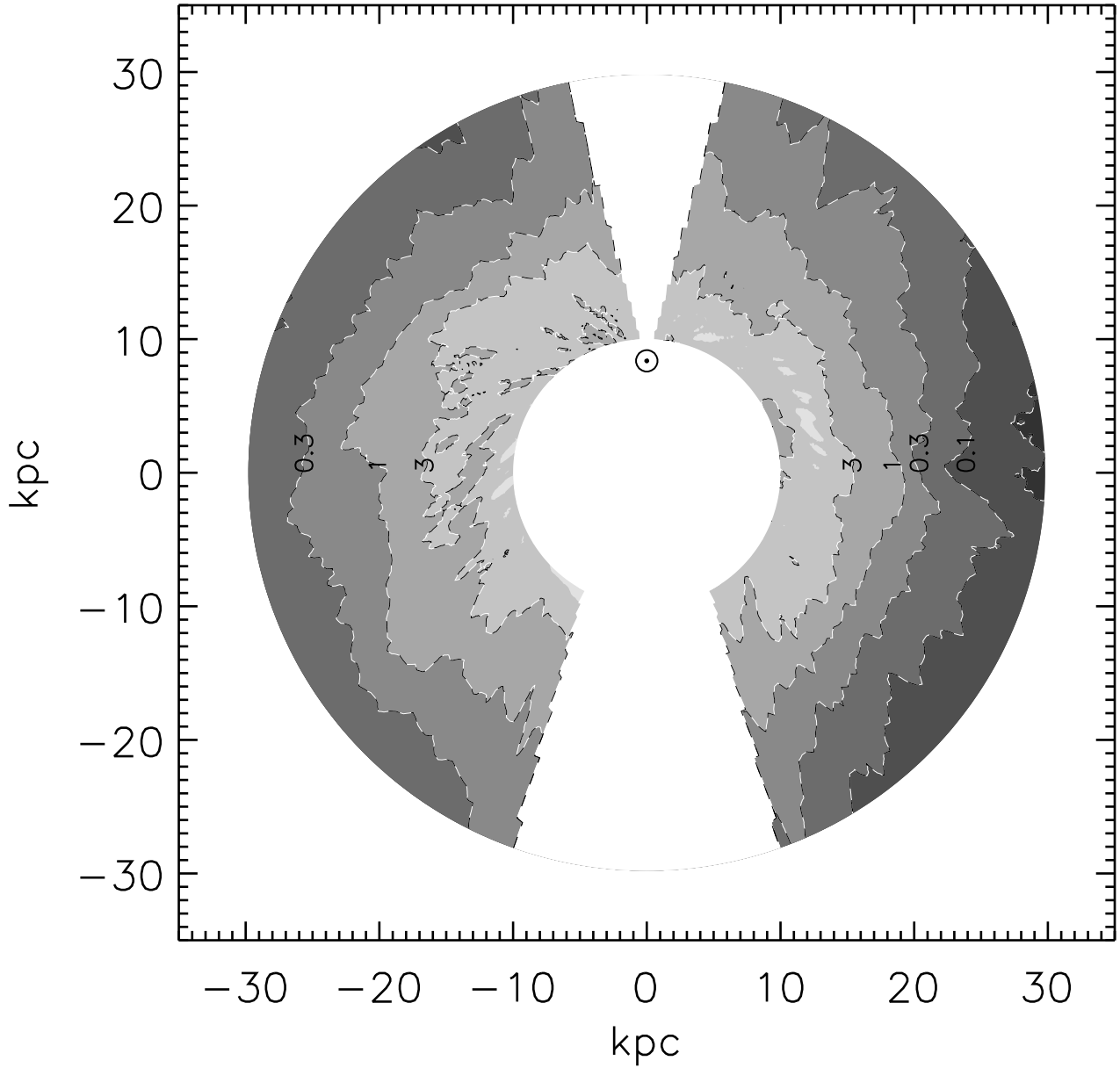}
\caption{\label{fig:sigma}A contour map of $\Sigma(R,\phi)$. Contours are evenly spaced in log$_{10}$ space, at $(3,1,0.3,0.1)~\mathrm{M}_\odot$/pc$^2$. The large blank regions directly towards and away from the Galactic center are the excluded regions for which it is hard to establish reliable distances. }
\end{figure*}

   We construct the Galactic disk surface density and mean height maps from $\rho_d(R,\phi,z)$. For example, we can sum the density over the $z$ dimension to construct the surface density associated with the disk  (Figure \ref{fig:sigma}):
   \begin{equation}
    \Sigma(R,\phi)=  \sum_{i=b}^t \rho_d(R,\phi,z_i) \Delta z
   \end{equation}
   The indices $t$ and $b$ represent  the top and bottom $z_i$ that have not been zeroed out by the dispersion filter. 
  The resulting figure clearly demonstrates the falloff of the disk surface density with radius. The contour lines are nicely continuous across the $\ell=0\degr$ and $\ell=180\degr$ lines because of our use of elliptical orbits; see the Appendix for a version of this figure without correcting for these orbits.  The jagged nature of the contours with $R\la 18$ kpc is  due in part to spiral arms \citep{LBH2006}. There is a region near $R\approx 27$ kpc and $\phi\approx90\degr$ with a smaller surface density than other regions at the same radius; this region has somewhat unusual features in all of our maps. 
   Also notice the excluded regions near the Sun-Galactic center line; these gaps will appear in all of our plots.
   
\begin{figure*}
\includegraphics[angle=90,scale=.75]{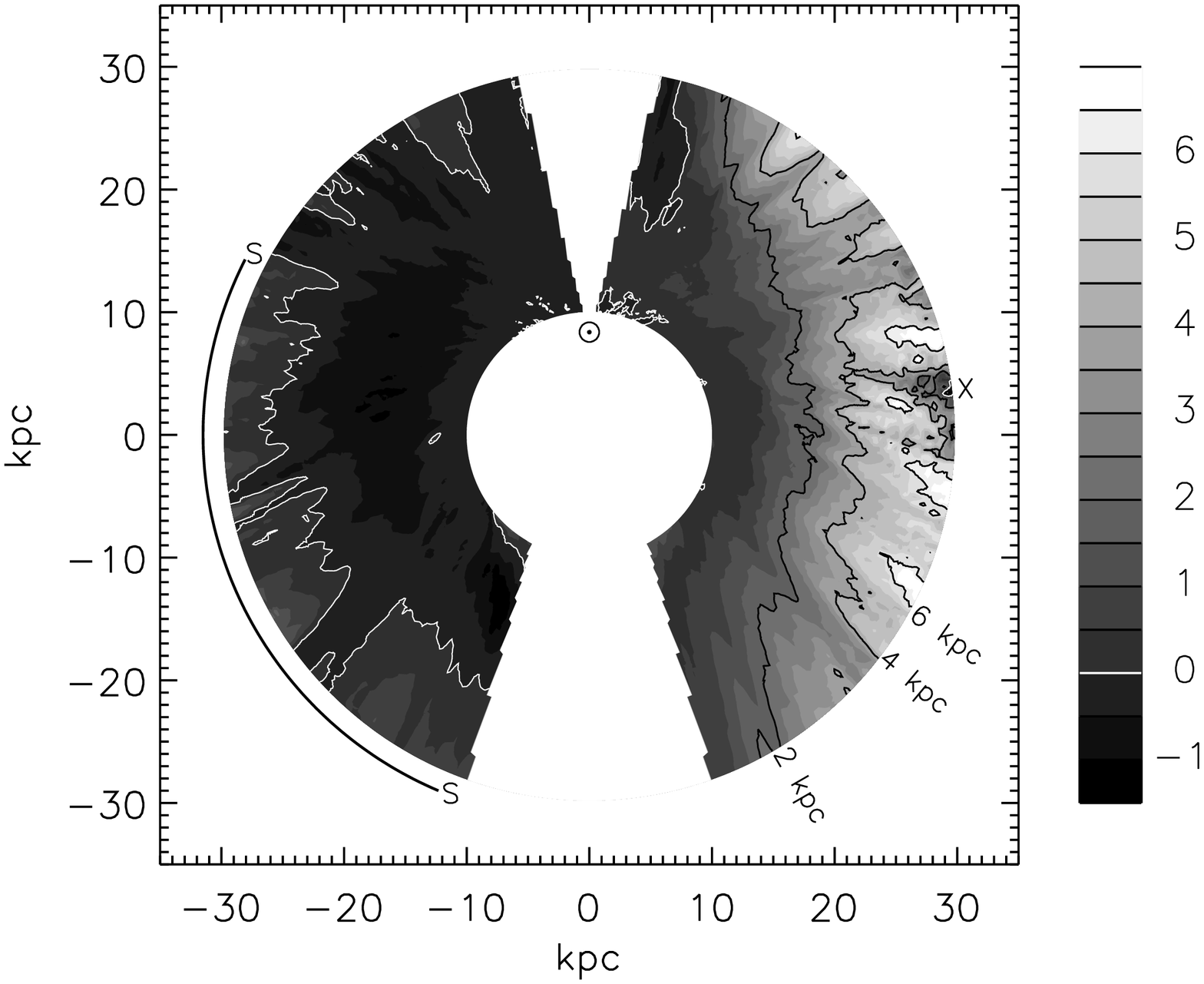}
\caption{\label{fig:height}A contour map of $h(R,\phi)$. The shading changes every 500 pc in vertical displacement with respect to the $b=0\degr$ plane.  A white contour is drawn at 0 pc; black contours are drawn at 2000, 4000, and 6000 pc. The colorbar on the right hand side is marked in vertical kpc. The two ``S'' labels mark the approximate extent of the scalloping as seen by previous authors. The ``X'' marks the sharp dip discussed in \S\ref{sec:wave} and \S\ref{sec:disc}. }
\end{figure*}

    We will be looking closely at the mass weighted mean height of the gas disk  (Figure \ref{fig:height}),
   \begin{equation}
 h(R,\phi)=\frac{1}{\Sigma(R,\phi)}\sum_{i=b}^{t} z_i \rho_d(R,\phi,z_i) \Delta z.
 \end{equation}
This height is calculated with respect to the Galactic midplane defined by $b=0\degr$. The Galactic warp is the most immediately evident feature in this map; the gas in the northern hemisphere peaks at $h\approx5$ kpc, while the southern gas descends only to $h\approx-1.5$ kpc,  consistent with previous maps of the Galaxy.  At least three vertical oscillations of magnitude $\approx 1$ kpc can be seen in the south from $\phi\approx-120\degr$ to $\phi\approx-20\degr$; these have previously been called the ``scalloping''. The approximate extent of the scalloping is marked with an arc connecting two ``S'' labels. The region of low surface density  noted in the discussion of Figure \ref{fig:sigma} ($R=27$ kpc, $\phi=90\degr$) has a height that seems anomalous when compared to surrounding gas (it is marked with an ``X''). Several features are elongated along lines of constant $\ell$ indicating some level of contamination by turbulent velocities and/or local gas. In the past these have been dubbed ``fingers of God'' because they all point back to the Sun.

In \S \ref{sec:warp} we will need a measure of the uncertainty in the mean height of each point in the map. We define this error using the sum of squared residuals, as is usual for least squares fits:
  \begin{equation}
  e^2(R,\phi)=\frac{\sum_{i=b}^{N=t}(z_i-h(R,\phi))^2 \rho_d(R,\phi,z_i) \Delta z}{(t-b+1)\Sigma(R,\phi)},
\end{equation}
where $t-b+1$ is the number of points in the calculation of the mean and surface density. This is only an approximation because it does not account for any of the uncertainty introduced in  $T_b(\ell,b,v_r)$, $v_\Pi(R)$, the interpolation to $\rho(R,\phi,z)$, or the dispersion filter. Typically, $e(R)/R\approx0.01$.

\begin{figure}
\includegraphics[angle=90,scale=.55]{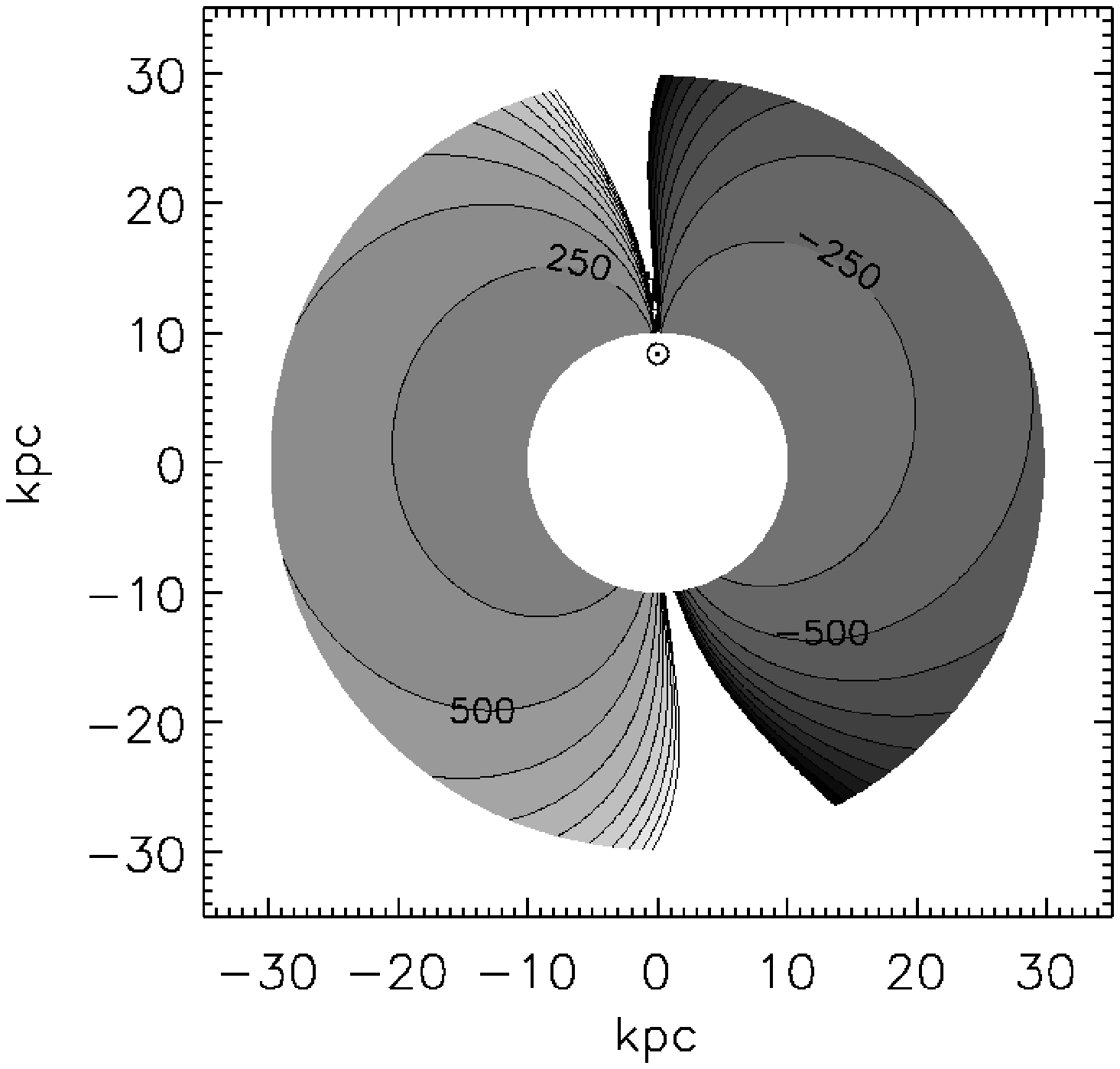}
\caption{\label{fig:uncert} A contour map of $\mathrm{d}r/\mathrm{d}v_r$, including the velocity contribution from the elliptical gas orbit correction. The axes are marked in kpc; the contours delineate regions of 250 pc/(km s$^{-1}$).  The contours rise close to the Galactic center and anticenter due to the small radial velocities of gas in those regions, demonstrating why the data in those directions are excluded from the rest of the study.  }
\end{figure}

Because of effects like turbulence, anomalous velocities, and spiral arms, the features in the mean height map may not correspond to the actual shape of the Galaxy. The severity of this effect can be seen by looking at a contour map of $\mathrm{d}r/\mathrm{d}v_r$ (see Figure \ref{fig:uncert}). While similar to previously published plots \citep{BT1986}, this figure also includes the correction for elliptical gas orbits. Given a turbulent velocity of magnitude $v_t$, features with coherent scales less than $v_t \mathrm{d}r/\mathrm{d}v_r$ could potentially be false signals. Furthermore, even real features will be blurred out or even incorrectly positioned over the same length scale.    A typical turbulent velocity is around 8 km s$^{-1}$. Small regions can differ from the flat rotation curve by 20-30 km s$^{-1}$ \citep{BB1993}. We do not expect that these distortions will significantly effect our Fourier analysis.   The steep contours near $\ell=0\degr$ and $\ell=180\degr$ demonstrate the need for the excluded regions, because small irregularities in velocity there result in large changes in distance.

\subsection{Thickness map}
  
We  also construct a measure of the  thickness of the disk. In contrast to the previous section, we  keep all of the complicated features of the HI in our calculation of the thickness. We do this because these features do contain information regarding  pressure and gravitational force, although they are a nuisance when calculating the mean height. In particular, our method for finding the thickness of the disk relies heavily on the tails of the vertical density distribution. Thus, in this section, we will work with $\rho$ instead of $\rho_d$.

\begin{figure*}
\includegraphics[angle=90,scale=.75]{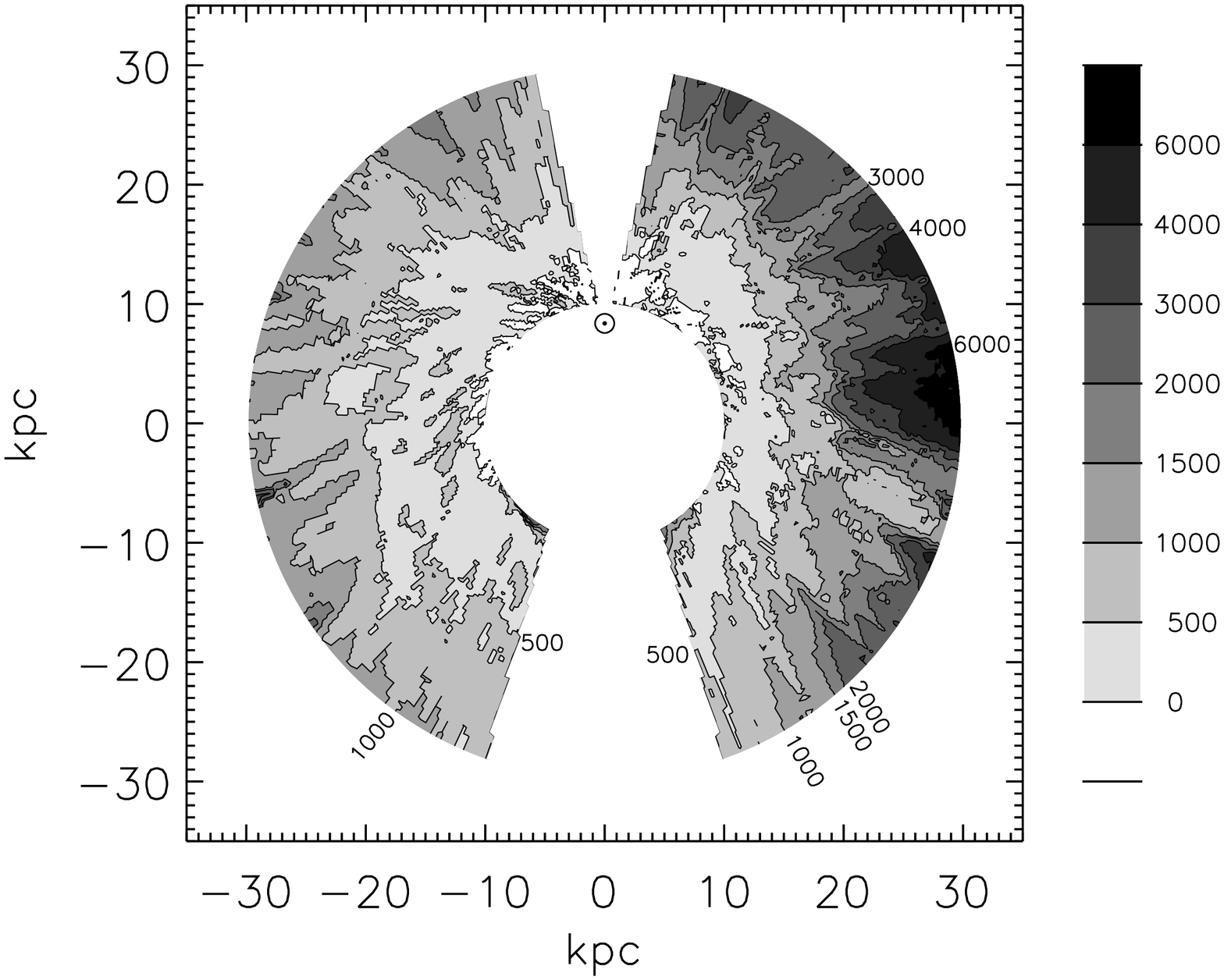}
\caption{\label{fig:disp} A contour map of $T_h(R,\phi)$; the thickness of the disk is approximately twice this quantity. Contours are drawn at 0, 500, 1000, 1500, 2000, 3000, 4000, and 6000 pc. The colorbar and the figure labels are in pc. }
\end{figure*}

Following \citet{HJK1982}, we define the first and third quartile points as $z_{j1}$ and $z_{j3}$ as the smallest and largest indices, respectively, that satisfy:
\begin{eqnarray}
\sum_{i=0}^{j1}\rho(R,\phi,z_i)\Delta z&\ge&M(R,\phi)/4\nonumber\\
\sum_{i=j3}^{141}\rho(R,\phi,z_i)\Delta z&\ge&M(R,\phi)/4.
\end{eqnarray}
We then define the half thickness (a factor of 2 smaller than \citet{HJK1982} to ease comparison with recent work): $T_h(R,\phi)=(z_{j3}-z_{j1})/2.$ Note that this implies that $T_h$ is quantized by $\Delta z/2$ (about 140 pc). This can lead to  inaccuracies in $T_h$ in places where the thickness is small, i.e. $R\approx10$ kpc. $T_h(R,\phi)$ is shown in Figure \ref{fig:disp}.  Some points near the sun actually have calculated half-thicknesses of zero; this is due to the comparatively small thickness of the disk in that region combined with our limited LAB survey range $|b|\le30\degr$ and poor grid resolution in $z$. 

The flaring of the disk with radius is immediately evident, as $T_h$ increases in magnitude by a factor of about 8 between $R\approx10$ and 30 kpc; this flaring was first seen in \citet{LK1963}. Asymmetry between the northern and southern halves of the disk is also prominent; the southern half of the Galaxy has a lower average thickness than the northern. This asymmetry was also evident in \citet{HJK1982} and \citet{BT1986}.
 The low surface density region around $R\approx27$ kpc has a very large thickness; the gas in the region has clearly been disturbed and dispersed by some mechanism. There  are regions of increased thickness close to $\ell=15\degr$ and $345\degr$ near $R\approx 10$ kpc probably caused by local gas.

\begin{figure}
\epsscale{1.2}
\plotone{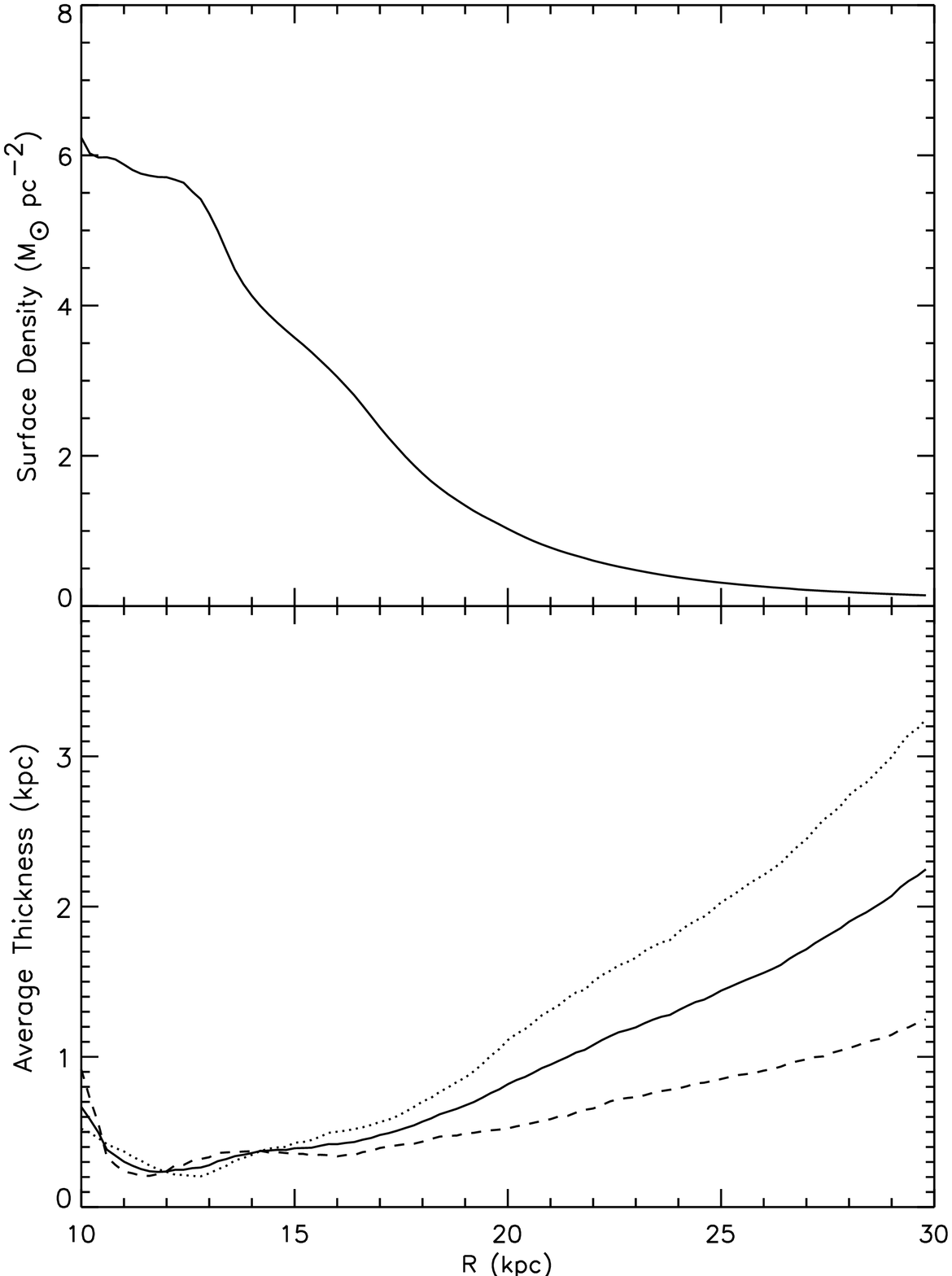}
\epsscale{1}
\caption{\label{fig:rfuncs}The upper panel is the disk surface density, and the lower is the half thickness of the gas as a function of radius. In the lower panel, the solid line is the average thickness over the whole disk, the dotted line is the average from  $0\degr<\phi <180\degr$, and the dashed line is from $180\degr<\phi <360\degr$. }
\end{figure}

Figure \ref{fig:rfuncs} plots the surface density and half-thickness of the HI layer averaged over $\phi$ as a function of $R$. This figure confirms our visual impressions from the surface density and dispersion maps by showing the falloff in the surface density and the flaring of the thickness with radius. The contamination by local gas described in the previous paragraph as well as the poor $z$ resolution problem mentioned at the beginning of the section account for the rise in the average of $T_h$ with decreasing $R$ near $R\approx 10$ kpc; there is no evidence the thickness of the disk actually behaves this way.
A gradient-expansion least-squares routine \citep{M1963} gives the best fit for the surface density beyond 14 kpc (where the exponential falloff begins) as: 
\begin{equation}
\Sigma(R)=4.5\times\exp[-(R-14~\mathrm{kpc})/ 4.3~ \mathrm{kpc}] ~\mathrm{M}_\odot~ \mathrm{pc}^{-2}.
\end{equation}

\section{Analysis}

\subsection{Global Structure}

Using the maps we have constructed, we conduct a quantitative investigation of the disk vertical structure. Since we are studying a disk, we  use a method that is  independent of rotation in $\phi$ and treats  $\phi =0$ and $\phi = 2\pi$ as the same point. Furthermore, the data are unevenly sampled due to the excluded regions within $15\degr$ of the Sun-Galactic center line.  We complete the analysis without extrapolating $h(R,\phi)$ in these regions, to avoid introducing any artifacts into the signal. Throughout this paper, we refer to different frequency oscillations in the disk. These frequencies will always be labeled by the number of oscillations they will complete in a full $2\pi$; thus the $m=1$ mode has a one maximum and one minimum in the disk.

\subsubsection{Galactic Warp}\label{sec:warp}

\begin{figure}
\includegraphics[angle=90,scale=.34]{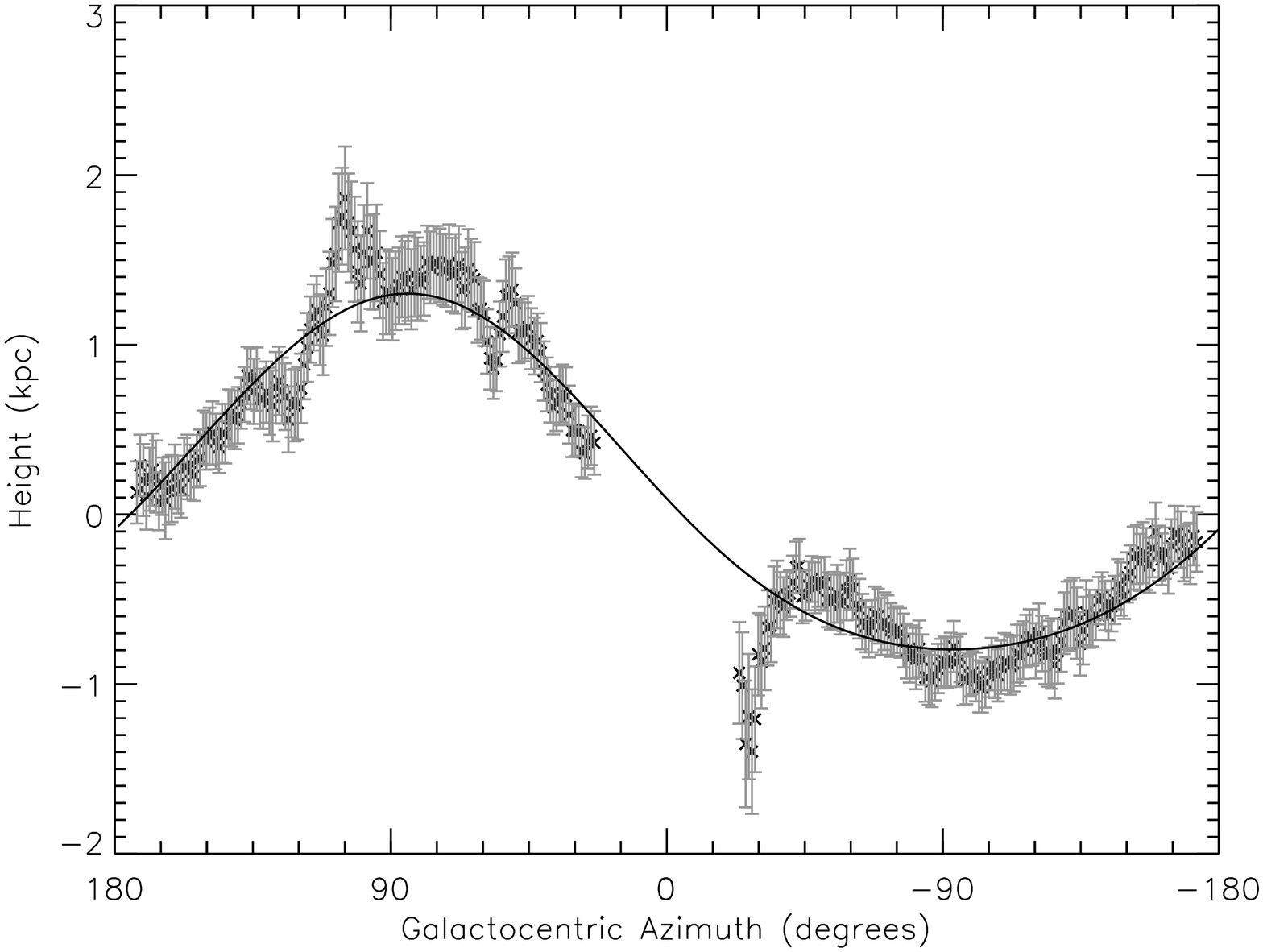}
\caption{\label{fig:warp16} The solid line is the warp fit $W(\phi)$ to $h$ at $R=16$ kpc. The approximate error in the mean height of the  disk at each point is represented by the error bars.}
\end{figure}

\begin{figure}
\includegraphics[angle=90,scale=.34]{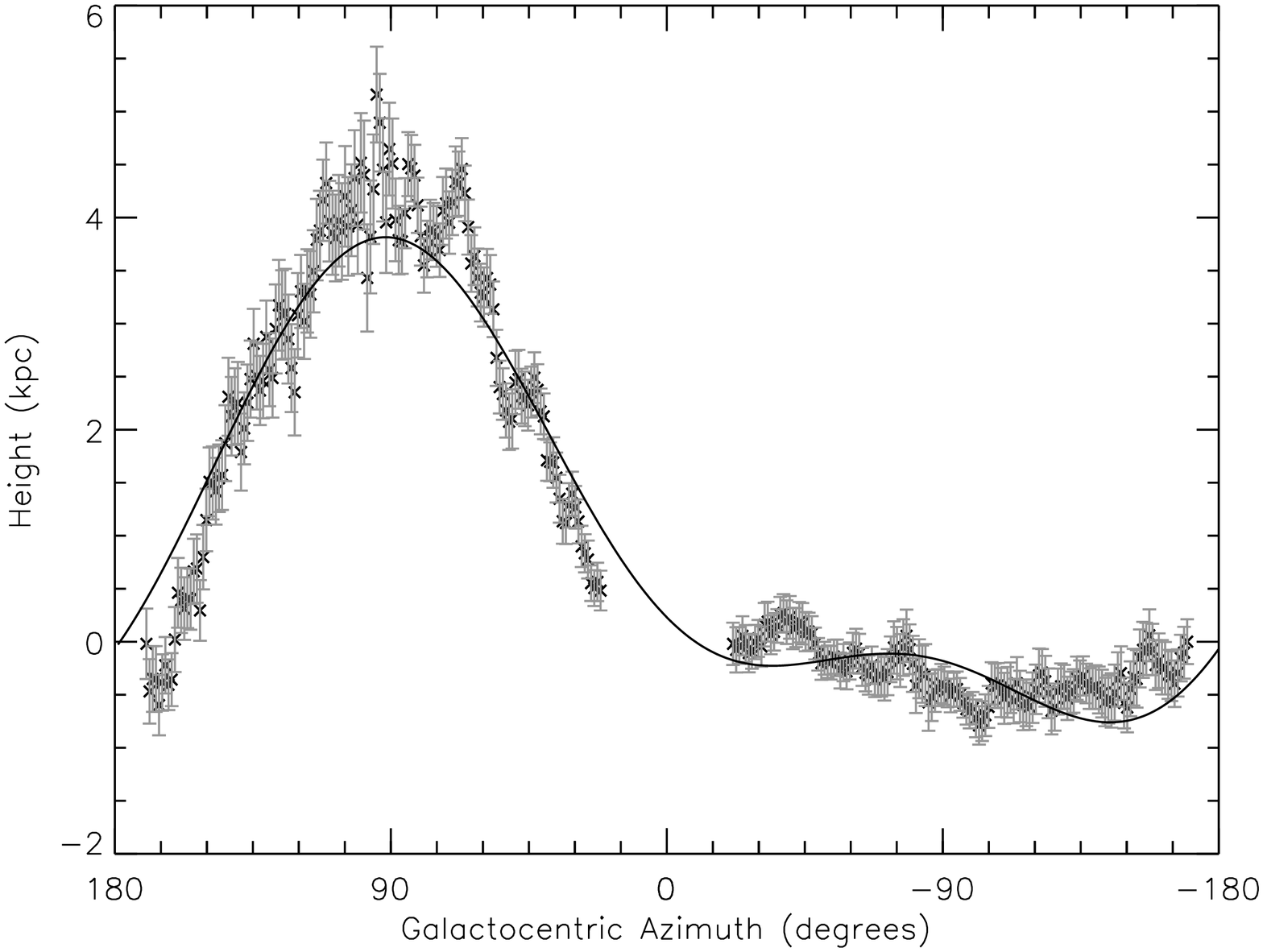}
\caption{\label{fig:warp22} The solid line is the warp fit $W(\phi)$ to $h$ at $R=22$ kpc. The approximate error in the mean height of the disk at each point is represented by the error bars.}
\end{figure}

\begin{figure}
\includegraphics[angle=90,scale=.34]{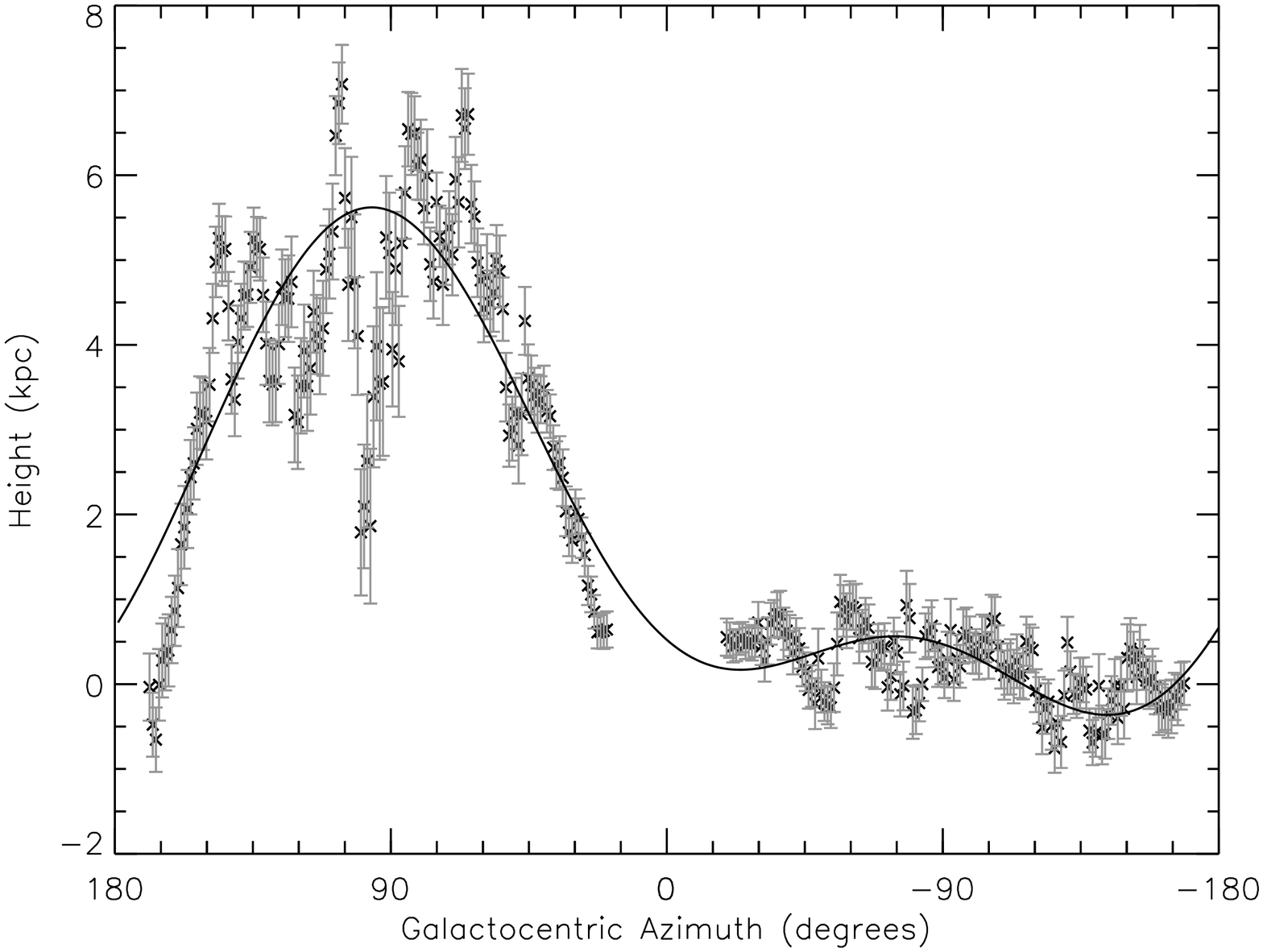}
\caption{\label{fig:warp28} The solid line is the warp fit $W(\phi)$ to $h$ at $R=28$ kpc. The approximate error in the mean height of the disk at each point is represented by the error bars.}
\end{figure}

The Galactic warp is the most prominent feature in Figure \ref{fig:height}.
A Lomb periodogram analysis of each radius ring (see \S \ref{sec:lomb}) reveals that the power in each of the $0,1,$ and 2 modes is consistently larger than that in any other mode for $R\ga 20$ kpc. At some radii $m=3$ is the next strongest mode, and at others it is $m=4$. 
Accordingly, we characterize the warp by an offset in the $z$ direction, plus two Fourier modes with frequency 1 and 2. We  fit each ring with the function:
\begin{equation}\label{eqn:fit}
W(\phi)=W_0 + W_1\sin(\phi -\phi_1)+W_2\sin(2\phi-\phi_2).
\end{equation}
Each of the three amplitudes $W_i$ and two phases $\phi_i$   in this fit is a function of radius, because we fit each radius ring independently.
We use the gradient-expansion  fitting algorithm  to perform this fit, weighting each point by the inverse of the squared estimate of the uncertainty in the mean height, $e^2(R,\phi)$.  The results of this fit for the rings at $R=16,22,$ and $28$ kpc are shown in Figures \ref{fig:warp16},\ref{fig:warp22}, and \ref{fig:warp28}. Error bars in these plots represent $e(R,\phi)$.  Each of these plots is a good fit; the offset and the two Fourier components are both necessary and sufficient to describe the large-scale structure of the disk. For illustrative purposes,  we will follow the ring at $R=28$ kpc through each step of the analysis.

\begin{figure}
\includegraphics[angle=90,scale=.34]{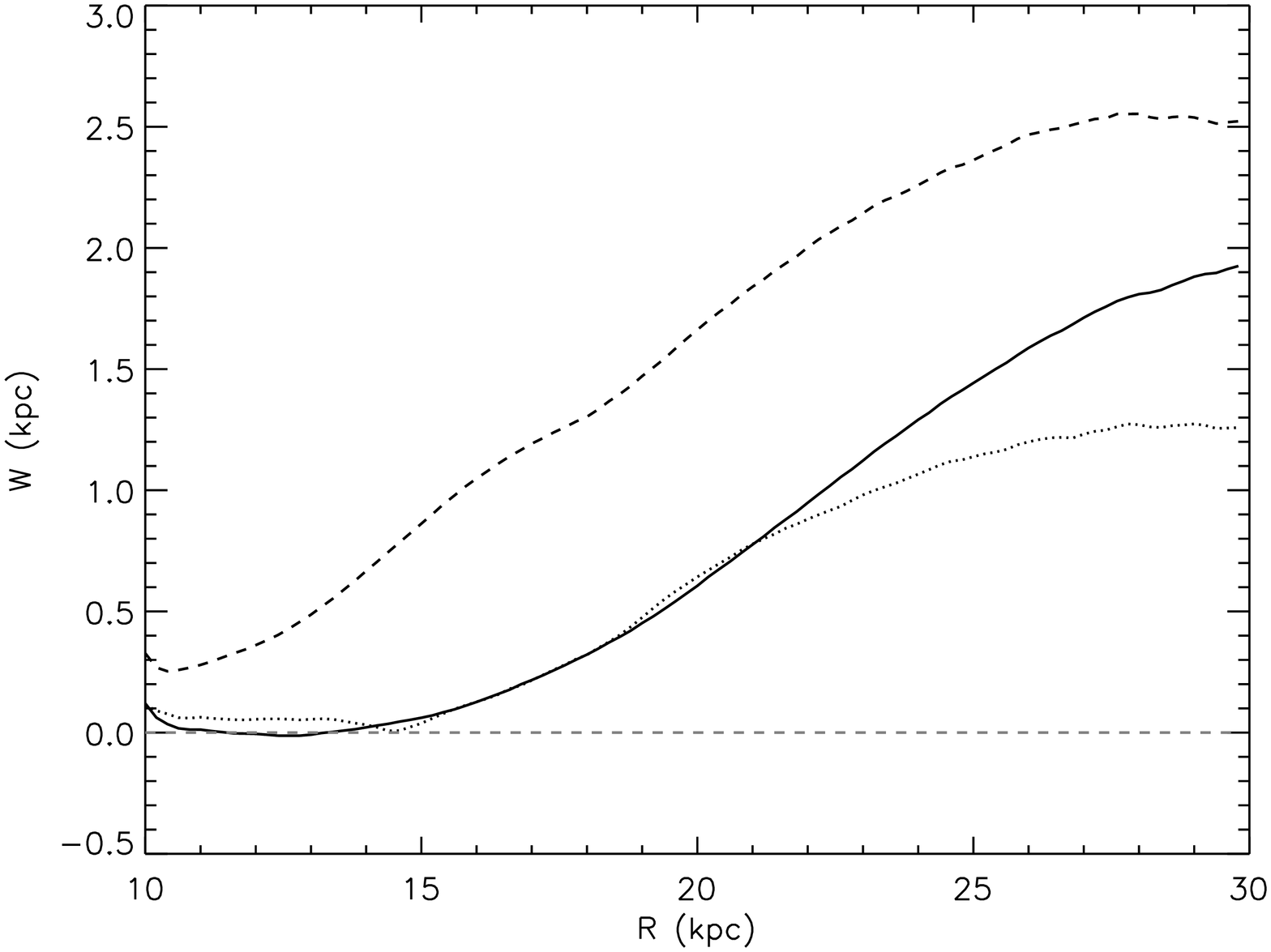}
\caption{\label{fig:warpamp} The evolution of the warp amplitude  parameters as a function of radius. $W_0$ is the solid line, $W_1$ is the dashed, and $W_2$ is the dotted. The dashed grey line is zero amplitude.}
\end{figure}

Following the evolution of the fit parameters at different places in the disk will tell us how the warp changes with radius.
The three amplitude parameters each increase monotonically, with the $m=0$ mode possibly reaching an asymptotic value near the far end of our radius range (Figure \ref{fig:warpamp}).  At $R\approx11$ kpc, the $m=1$ mode dominates the shape of the warp; the other two modes do not become important until $R\approx 15$ kpc. This plot implies that the $m=1$ mode has power even at the edge of our grid, thus we cannot establish the onset of the warp. We do a linear least-squares fit on the growth of each warp parameter using the function 
\begin{equation}\label{eqn:warpfit}
W_n=k_0+k_1\left(R-R_k\right)+k_2\left(R-R_k\right)^2
\end{equation}
where  only points at $R_k$ and beyond are weighted in the fit. $R_k$ is arbitrarily chosen to be near where each of the three modes starts growing. The value of $k_0$ for each fit is strongly correlated to the choice of $R_k$.

\begin{deluxetable}{rrrrr}
\tablecaption{\label{tab:warpfit} Parameters resulting from a linear least-squares fit to the warp}
\tablehead{
\colhead{$m$}&\colhead{$R_k$ (kpc)} &\colhead{$k_0$ (pc)}&\colhead{$k_1$ (pc kpc$^{-1}$)}&\colhead{$k_2$ (pc kpc$^{-2}$)}}
\startdata
0&15&-66&150&-0.47\\
1&10&9&197&-3.1\\
2&15&-70&171&-5.3
\enddata
\end{deluxetable}

\begin{figure}
\includegraphics[angle=0,scale=.44]{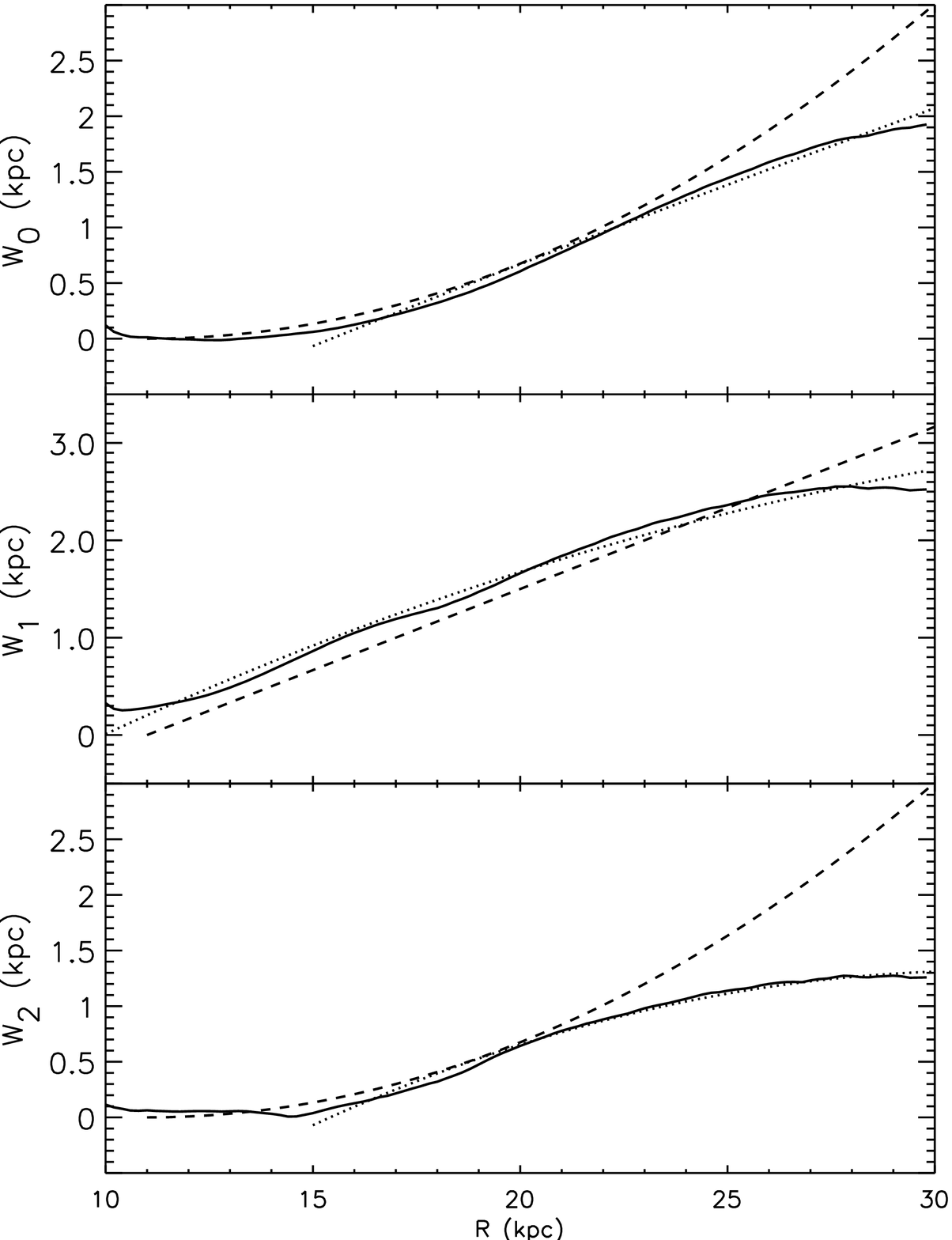}
\caption{\label{fig:binney} In each panel, the solid line is the calculated amplitude of the the mode from (\ref{eqn:fit}), the dashed line is the fit from BM, and the dotted line is the fit described in (\ref{eqn:warpfit}). Top panel: $m=0$. Middle panel: $m=1$. Bottom panel: $m=2$.}
\end{figure}

\citet{BM1998} (hereafter BM) discuss an approximation to the warp that is of similar form to our fit. They also fit the first three modes, but they fix the line of zeros to lie along the Sun-Galactic center line. In our fit, this would be equivalent to setting $\phi_1$ and $\phi_2$ to zero. Also, BM fix $W_0$ and $W_2$ to be the same. Note that our warp data are adjusted for elliptical gas orbits, and are filtered according to the procedure described in \S\ref{sec:proc}, whereas BM's data are not. Fig.~\ref{fig:binney} compares the mode amplitudes calculated from the data, the fit from BM, and our fit. BM overestimate the strength of the warp starting at around $R\approx22$ kpc for the $m=2$ mode, 24 kpc for the $m=0$, and 27 kpc for the $m=1$. However, the BM fit matches our data fairly well for the radii where the warp is growing most rapidly.

\begin{figure*}
\includegraphics[angle=0,scale=.95]{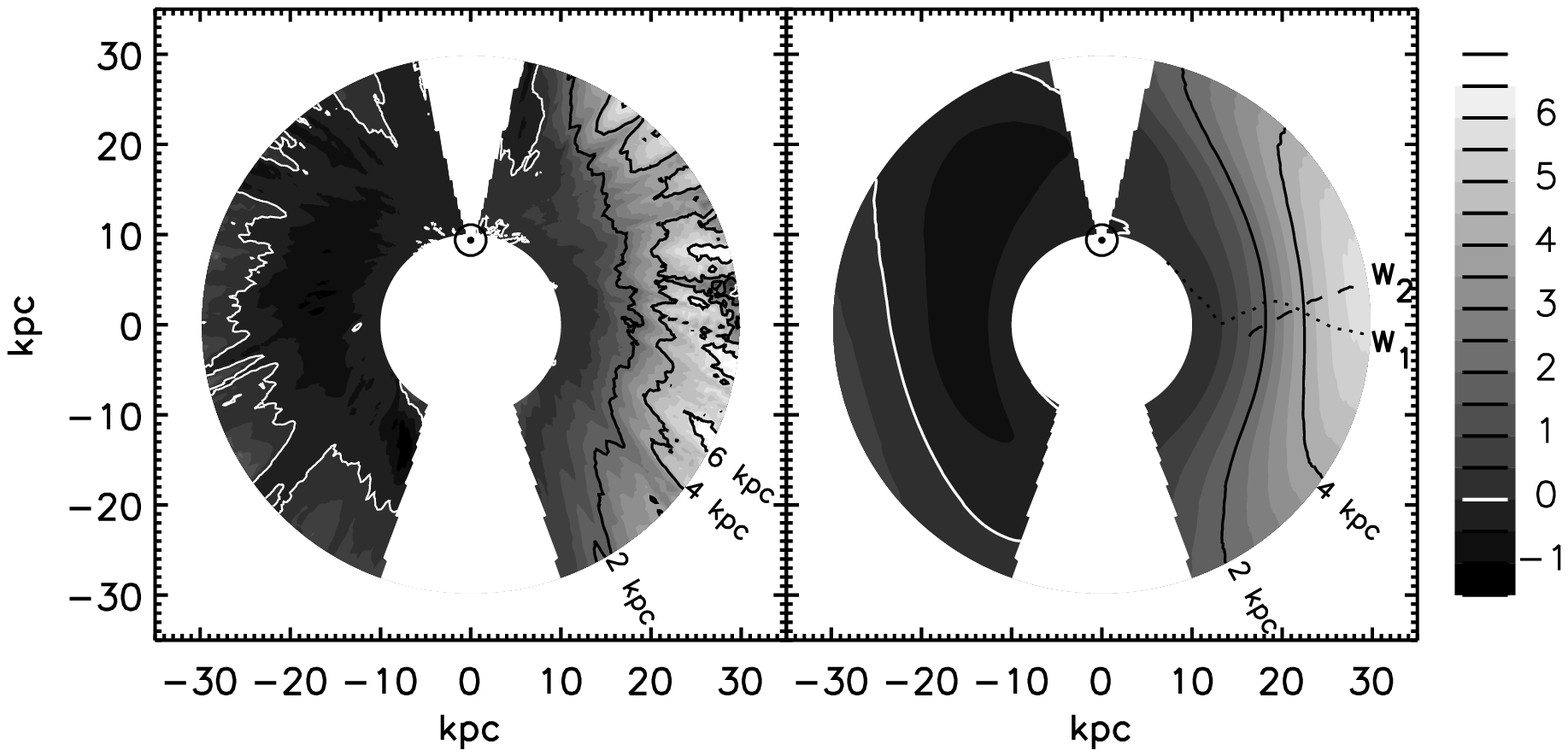}
\caption{\label{fig:phase} Left panel: $h(R,\phi)$. Right panel: The fit to the warp is plotted, along with the lines of maximum amplitude for the $m=1$ (dotted) and 2 (dashed) modes. These lines are marked $W_1$ and $W_2$ respectively. The white contour line denotes a height of 0 pc; black lines mark the 2000, 4000, and 6000 pc elevations. }
\end{figure*}

The line of maxima for the $m=1$ mode and one of the two lines of maxima for the $m=2$ mode are marked on the contour map of the warp fit (Figure \ref{fig:phase}). Since $\phi_1$ and $\phi_2$ are actually the line of zeros, the lines of maxima are shifted $90\degr$ and $45\degr$ from these values in our fit, respectively. The parameter $\phi_2$ is not well determined at small radii in our fit because the amplitude of the $m=2$ Fourier component is very small in that region.  For this reason, we do not plot $\phi_2$ in the region where $W_2$ is less than 150 pc. There is little evidence for precession in the lines of maxima for the two modes, and the line of maxima for the $m=1$ mode is roughly aligned with one of the lines of maxima of the $m=2$ mode; for each radius their difference in $\phi$ is less than $12\degr$. \citet{B1988} has examined how the two  lines of zeros for the mean height change in $\phi$ as a function of radius; they appear to stay roughly aligned with the Sun-Galactic center line as $R$ changes from $R_0$ to 26 kpc. Fig.~\ref{fig:phase} shows that the line of zeros for the sum of our three warp modes falls within the excluded region, consistent with the earlier work.

The three component fit shown in Figure \ref{fig:phase} does a good job of reproducing the large-scale features in the mean height map. This is not surprising, given that the $m=0,1,$ and 2 modes are the strongest in the Lomb periodogram; differences between the two plots are due to power in high frequency modes. We will now examine the differences between the mean height map and the warp fit.

\subsubsection{Scalloping}\label{sec:lomb}

From the fit to the warp, we can determine a function for the scalloping: $s(R,\phi)=h(R,\phi)-W(R,\phi)$. We continue to  assume no information regarding the shape of the disk in the excluded regions. 
We can now look for higher $m$ modes of oscillation that may be present in $s(R,\phi)$. Such a signal would be present if the gas were coherently moving up and down as a function of $\phi$. The Lomb periodogram is a  useful numerical method for detecting periodic signals in unevenly spaced data \citep{NumRec}. It provides the same amplitude we would determine by using a linear least squares fit, even if the signal is not sinusoidal, while allowing for a straightforward error analysis. The data are weighted equally per point, which is necessary to deal with unevenly sampled data. To ease comparison with the warp component amplitudes in Fig.~\ref{fig:warpamp}, we use an unnormalized Lomb periodogram and take the square root of the power to get the amplitude.

We use a Monte Carlo algorithm to determine the noise level for the Lomb periodogram. For each ring, the null hypothesis is  Gaussian white noise with the same dispersion as the data in $s(R)$. We construct $10^3$ sets of noise for each ring, run the Lomb periodogram on each set, and record the highest peak in the Lomb amplitude. We then determine the distribution of peaks in these amplitudes, and define the 95\% confidence interval as the amplitude just larger than 95\% of the noise amplitude peaks. 
We use the same technique to determine the 99\% confidence level.   To conclude that a signal is real and not caused by noise, the signal strength must cross these confidence levels.
We found nearly identical noise levels by scrambling the order of the data instead of using Gaussian white noise, indicating that these calculations are robust.

\begin{figure}
\includegraphics[angle=90,scale=.34]{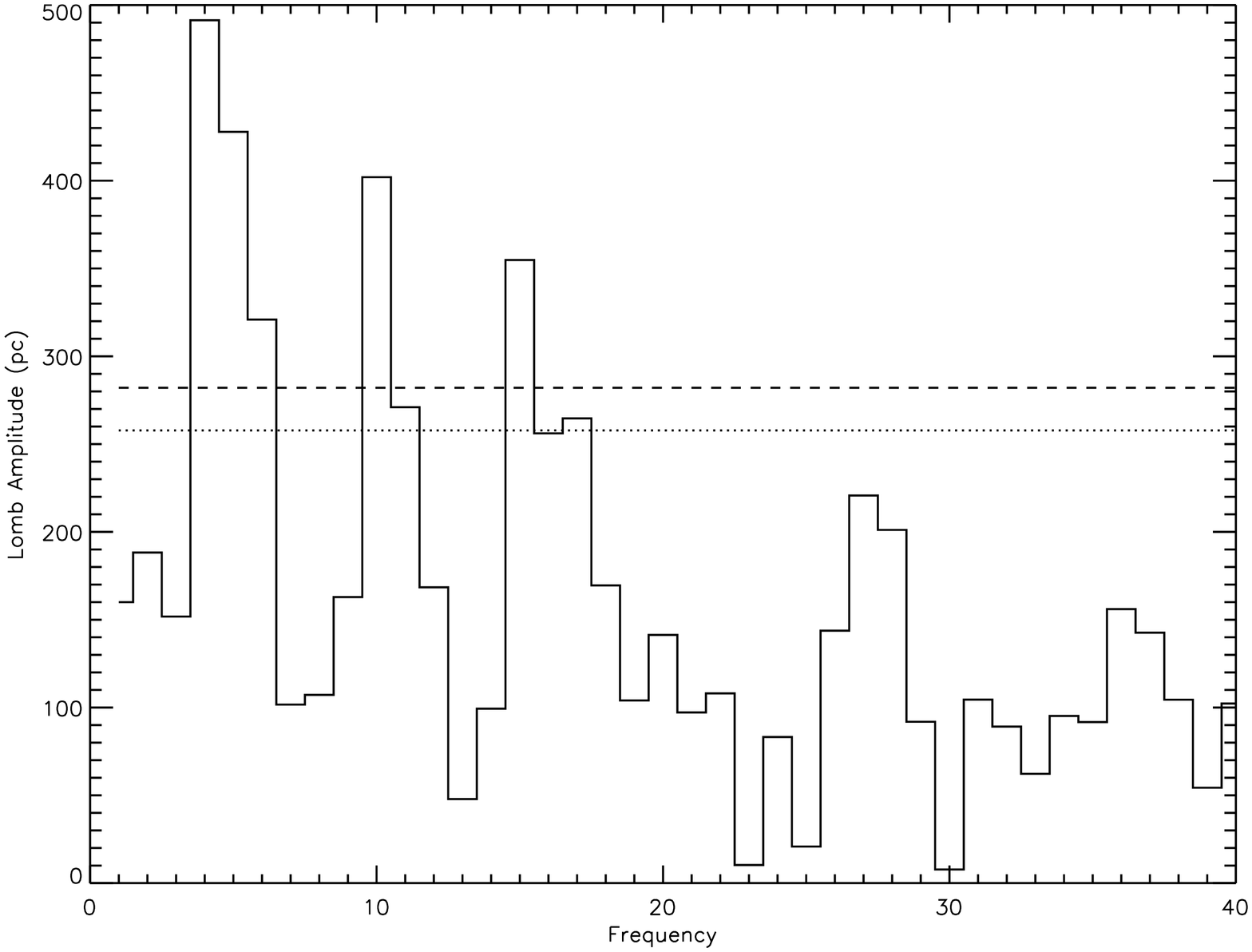}
\caption{\label{fig:lombring28}The Lomb Periodogram for the ring at 28 kpc, once the warp has been removed. The 95\% and 99\% significance levels are drawn as dotted and dashed lines, respectively.}
\end{figure}

The Lomb periodogram of $s(R,\phi)$ for the $R=28$ kpc ring is shown in Figure \ref{fig:lombring28}. 95\% and 99\% confidence levels are marked as thresholds in amplitude. At this radius, there is significant strength in modes 4-6, 10, and 15. These modes are markedly weaker than the warp components at the same radii, which have amplitudes of 1-2 kpc.

In the case of unevenly spaced or missing data points, simple sine waves are not eigenmodes of the system.  One way to see this is to take a Lomb periodogram of a pure sine wave on our $\phi$ grid with missing data; a small amount of the power will leak into other frequencies. Thus subtracting out the warp from $h(R,\phi)$ influences the amplitudes of the higher order modes from the Lomb periodogram because individual $m$ modes are not independent.   We argue that the subtraction is nonetheless acceptable because the warp and the scalloping appear to be due to physically distinct phenomena; studying them is much easier once they have been separated. Removing the strongest modes will also result in a more accurate power spectrum of the weaker frequencies, since we eliminate the power leakage from the stronger to the weaker modes.

\begin{figure*}
\epsscale{1.15}
\plotone{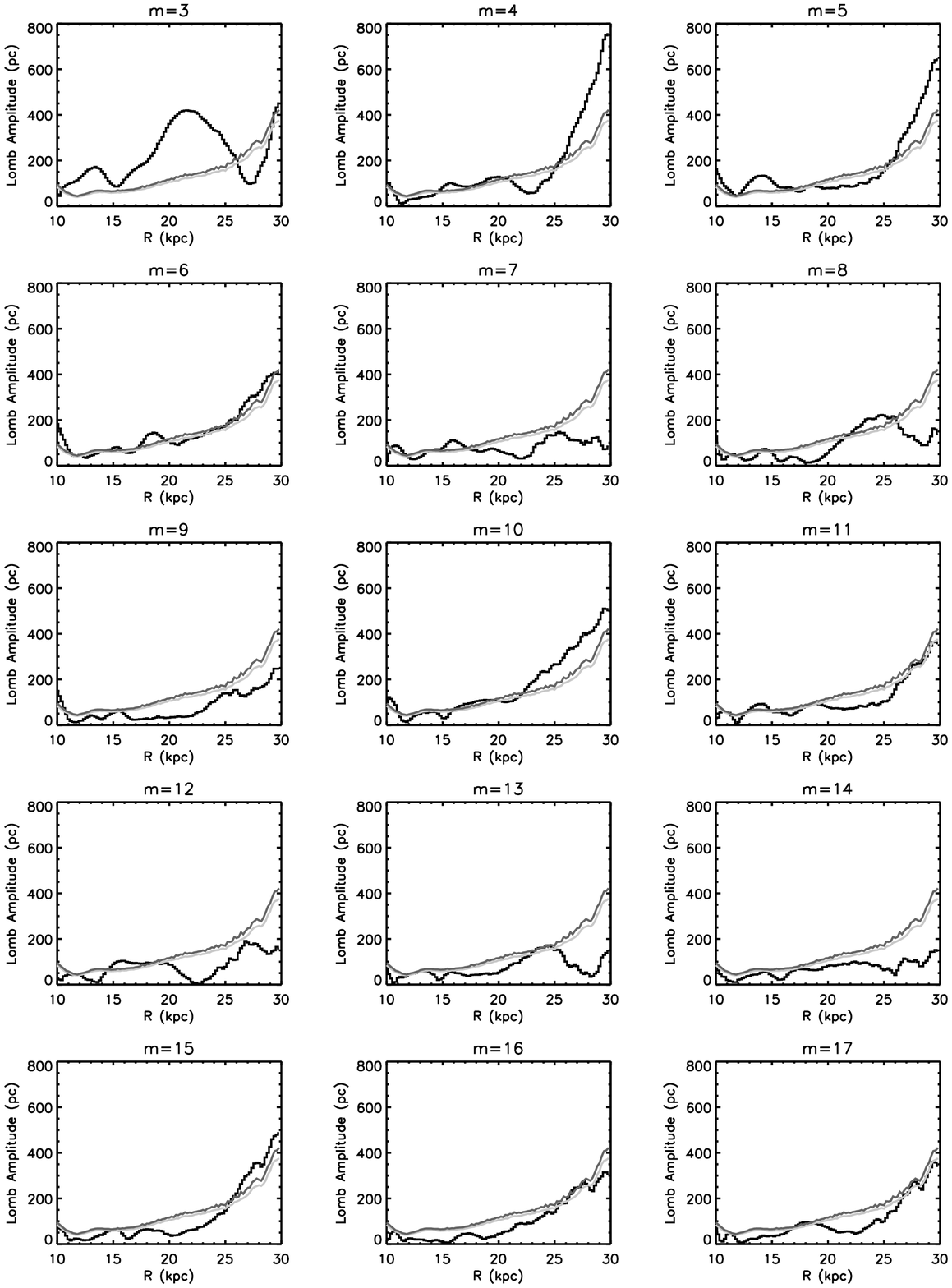}
\caption{\label{fig:lombfreqs}The Lomb periodogram amplitudes as a function of radius, once the warp has been removed. The 95\% and 99\% significance levels are drawn as light grey and dark grey lines, respectively.}
\epsscale{1}
\end{figure*}

 Figure \ref{fig:lombfreqs} shows how the amplitudes in  several different $m$ modes evolve with $R$.  Modes 4-6, 10, and 15 appear to increase in strength at outer radii, while mode 3 is strongest in the intermediate radii in our map. Other modes have no detectable strength because they do not cross the 95\% confidence level at any radii. Also notice how the confidence thresholds rise with radius because of the increase in the dispersion of $s(R,\phi)$.

Another consequence of using the Lomb periodogram analysis on unevenly spaced data is that it is possible to be fooled by a false signal due to ringing from interference between different $m$ modes.  It is difficult to protect against this, but if two  strong modes were interfering with each other to produce a third signal, we would expect at least two of the signals to increase in strength at the same radius. The strongest modes in Figure \ref{fig:lombfreqs} become significant at different radii and have visually different evolution with radius; we conclude that these modes are real.

\subsection{Local Structure}\label{sec:wave}

  The Lomb periodogram cannot be used to study the local structure of the disk since it cannot determine where in azimuth each mode is strong. Imagine a situation where, like a falling stone creating ripples on the surface of a pond, something passes through the HI disk and excites a local series of vertical oscillations. The oscillation will add power to some frequencies in the Lomb periodogram, but this effect may be dwarfed by oscillations elsewhere in the disk. We wish to detect these sorts of perturbations and learn where in the disk they are prominent.

One way to draw out this type of structure is to use a wavelet analysis. Wavelets are ideal for our purposes because they are localized in both position and frequency space; in other words, they will show where in the disk a given mode of oscillation is dominant. In wavelet theory, it is beneficial to choose a mother wavelet that is similar in shape to the signals you are searching for. Since we are looking for sinusoidal perturbations, we use the normalized Morlet wavelet with $\omega_0=6$; this is simply a sine wave times a Gaussian envelope. In the same way that a Fourier transform breaks down the frequency structure of a signal using sines and cosines, a wavelet transform breaks down a signal in terms of a set of these Morlet wavelets centered at different spatial positions and with a range of frequencies. However, in a continuous wavelet transform like the one we perform, the different wavelet functions are not orthogonal.

To avoid having to interpolate or zero out the excluded regions, we will examine the northern and southern halves of the Galaxy separately. As in \S \ref{sec:lomb}, we will work with the mean height function once the warp has been subtracted, $s(R,\phi)$.
Points in frequency and position space that would involve the excluded regions are defined as being inside the  ``cone of influence''; points in this region are subject to edge effects, and are therefore discarded. The points affected by this are not just those in the excluded regions but also those adjacent to these regions, within some range set by the wavelength (and thus the frequency) of the particular mode \citep{TC1998}. Thus, lower frequency modes will have a larger portion of the disk fall inside the cone of influence, and have to discard a larger range of points.

It is also important to have some analytic measure of which peaks in the filtered power spectrum  are significant. Significance levels are discussed in detail in \citet{TC1998}. We construct a combination of parameters that have a $\chi^2$ distribution, and count as significant those that cross the 95\% confidence level threshold.
Again, we use Gaussian white noise to model the randomness in the height function. 

\begin{figure}
\epsscale{1.15}
\plotone{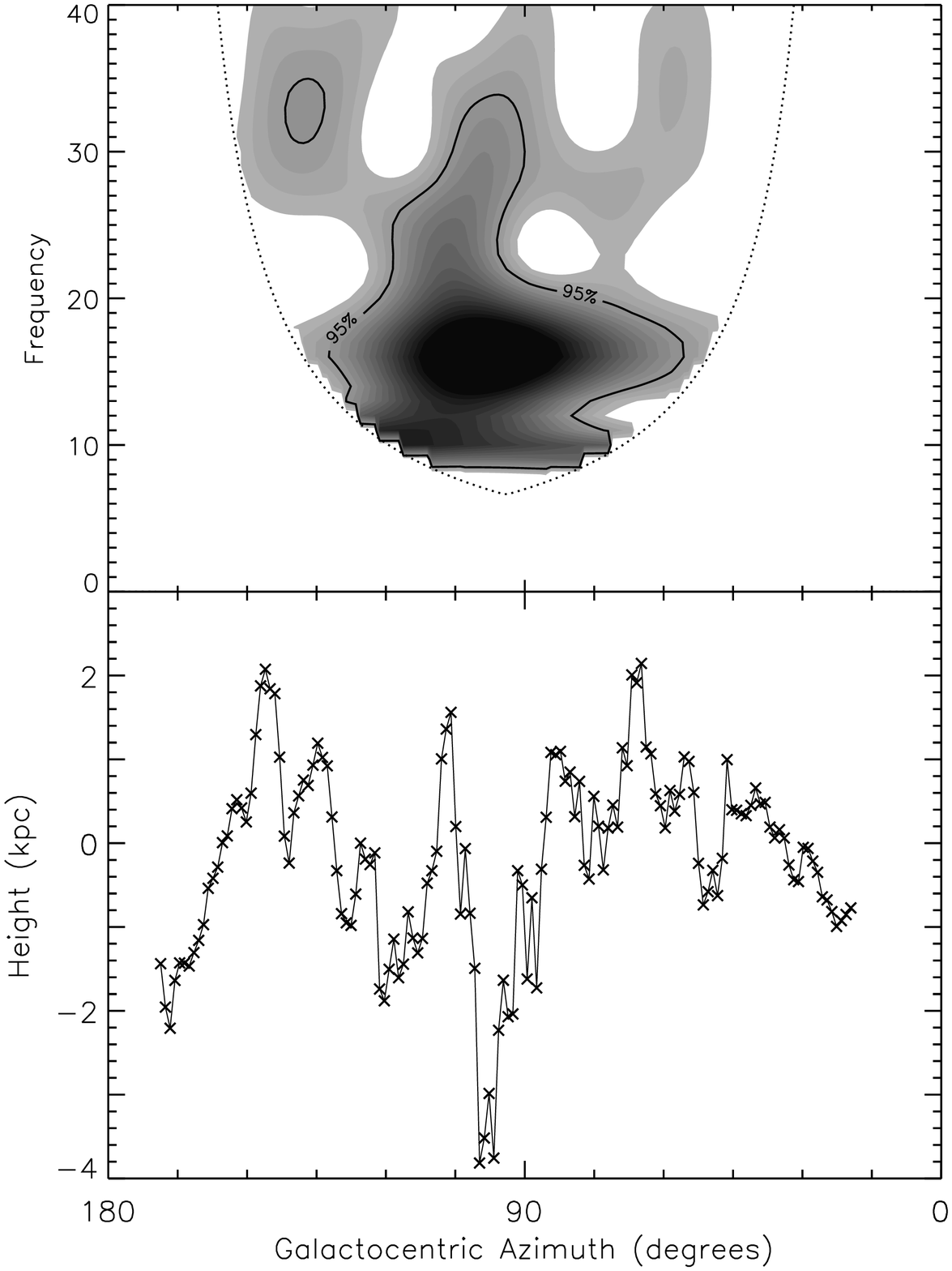}
\caption{\label{fig:specring}The filtered wavelet power spectrum for the northern half of the $R=28$ kpc ring once the warp has been subtracted. The 95\% significance level is marked with a solid line, and the cone of influence is outlined with a dotted line. The function $s(28~\mathrm{kpc},\phi)$ is plotted in the lower panel.}
\epsscale{1}
\end{figure}

The wavelet power spectrum, $\overline{W_n}^2$, for the northern half of the $R=28$ kpc ring is plotted in Figure \ref{fig:specring}; the southern half is shown in Figure \ref{fig:specring2}. The mathematical details of the wavelet transform are summarized in the Appendix.
This ring has several regions with significant power. These regions have a large width in both dimensions because wavelets do not have precise resolution in either position or frequency space. Much like the uncertainty principle, the cost of using a technique that gives both position and frequency information is mediocre resolution in both.  The power around $\phi\approx90\degr$ comes from the sharp dip at that azimuth; as the figure shows, this causes ringing for a large range of frequencies. The power at $-45\degr\ga\phi\ga-90\degr$ is due to the ``scalloping'' in that region. 

\begin{figure*}
\includegraphics[angle=90,scale=.9]{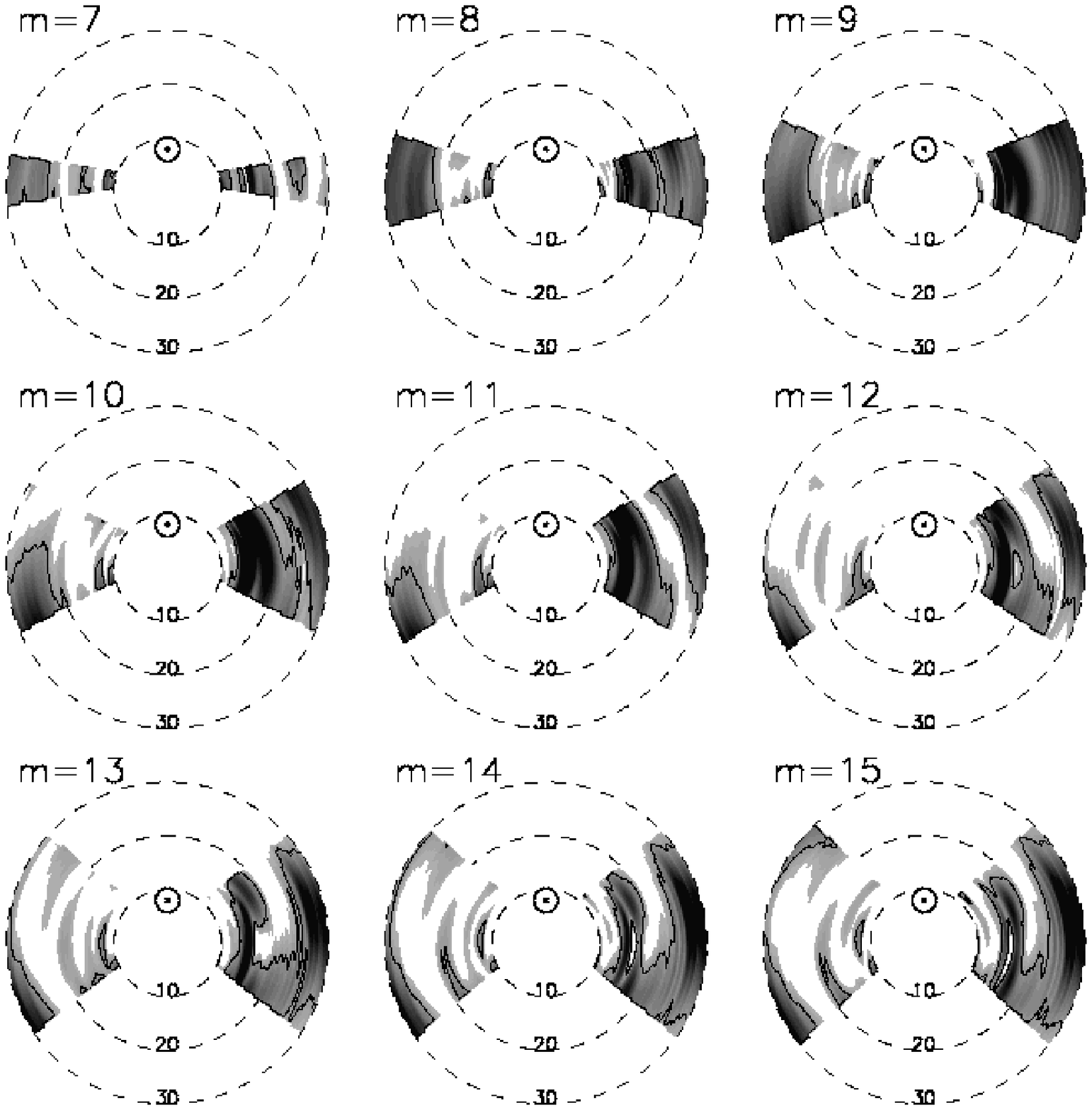}
\caption{\label{fig:waveplot}The filtered wavelet power spectrum for the outer Galaxy. Darker contours represent higher power, and the 95\% significance level is marked with a solid line. The labels on the rings give the distance from the Galactic center in kpc.}
\end{figure*}

The filtered wavelet power spectra for several different bands are plotted in Figure \ref{fig:waveplot}. The band that is labeled $m=7$ is actually the filtered sum of the power spectra that satisfy $6.5\le m \le 7.5$, and so on.  Note that since we calculate the power spectra for the northern and southern halves independently, the significance contours are different for the two hemispheres, in addition to being a function of $R$. In practice, this occurs because the variance of $s(\phi)$ for the northern half of the Galaxy is larger than that for the southern, and the strength required of the power spectrum  to cross the 95\% confidence level is directly proportional to the variance.

\begin{figure}
\epsscale{1.15}
\plotone{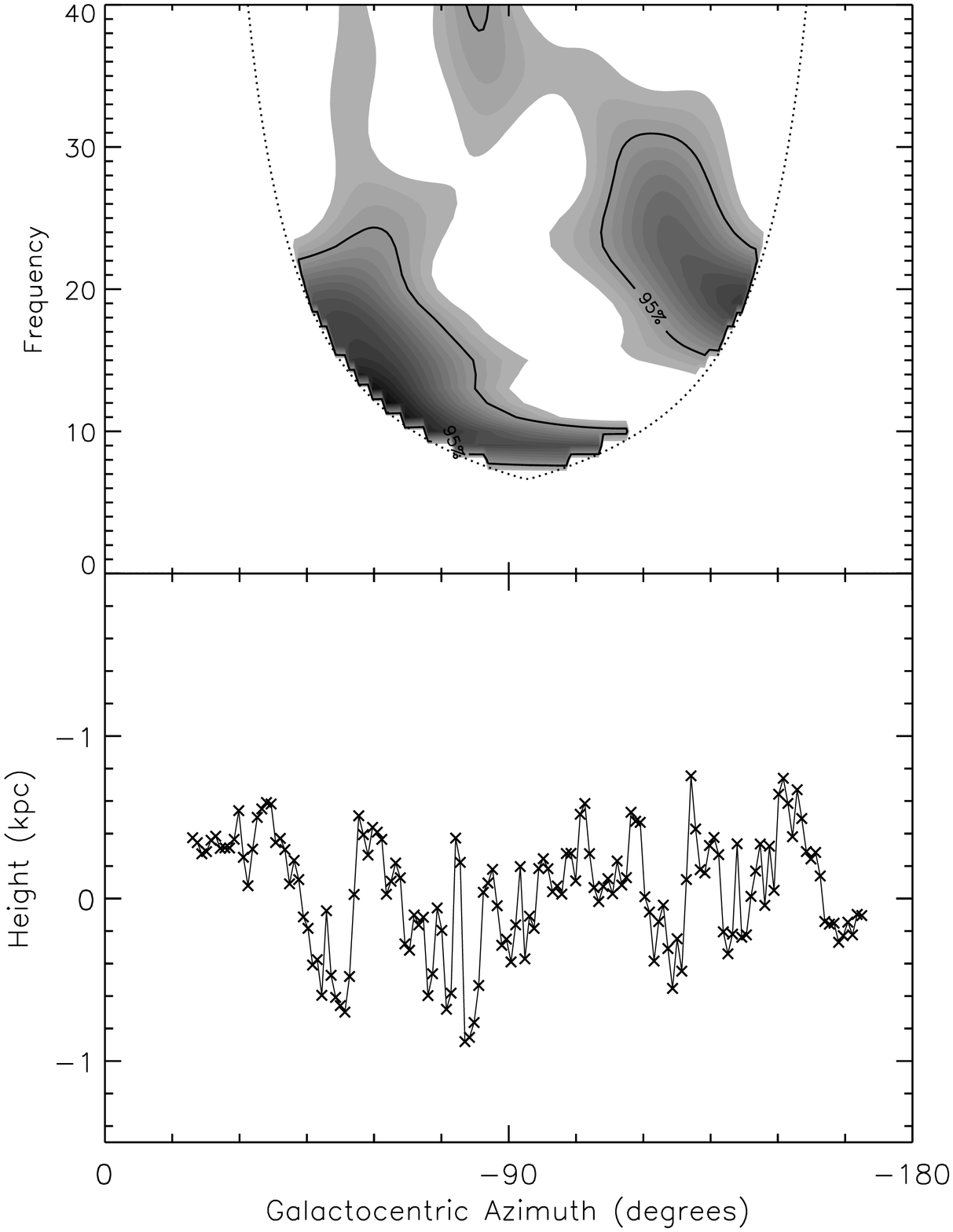}
\caption{\label{fig:specring2}The filtered wavelet power spectrum for the southern half of the $R=28$ kpc ring once the warp has been subtracted. The 95\% significance level is marked with a solid line, and the cone of influence is outlined with a dotted line. The function $s(28~\mathrm{kpc},\phi)$ is plotted in the lower panel. }
\epsscale{1}
\end{figure}

\section{Discussion}\label{sec:disc}

 We grouped the $m=0,1$ and 2 modes together in the warp because of their similar magnitudes in a Lomb periodogram. A close look at their dependence with $R$ (Fig.~\ref{fig:warpamp}) demonstrates that there are additional similarities.  Both the 0 and the 2 mode are near zero until about 15 kpc from the Galactic Center.  The 0 mode grows  linearly from this point outwards while the 2 mode
  declines a small amount, and then grows approximately linearly by about 1.2 kpc. In comparison, the $m=1$ mode starts out fairly large ($\approx 300$ pc), declines slightly and then grows by about an order of magnitude.  Though the $m=2$ mode does break off from this pattern at larger radii, all three components  grow essentially monotonically, approximately linearly, and with similar slopes  over the range  $15\la R\la 22$ kpc. This may be a clue that their origins involve the same physics, and helps to justify our classification of higher frequency modes as scalloping. Studying the radial dependence of the warp inside the solar circle seems a fruitful way to learn more about these three components.
  
The filtered wavelet power spectrum maps (Fig.~\ref{fig:waveplot}) are  a representation of the scalloping in the outer Galaxy. Due to frequency-position uncertainty relations, it is impossible to establish precisely the oscillation frequency of any local disturbance. This uncertainty manifests itself in the wavelet transform by perturbations that are only somewhat localized in frequency and position space, and therefore have some width in both.

These maps demonstrate that the $m\approx10$ scalloping found near $\ell\approx 310\degr$ and $R\approx 25$ kpc mentioned in previous work \citep{HJK1982,KBH1982} is real. The strong power around $m=10$ in the wavelet transform is accompanied by a large amplitude  in the Lomb periodogram for $m=10$ at the corresponding radii. 

Wavelet transforms are also subject to the same problems that bedevil traditional Fourier transform approaches. For example, there is a large amount of power around $R\approx30$ kpc and $\phi\approx90\degr$ in $8\le m\le15$. This power is most likely not a result of scalloping in all of these different frequencies over some range of $\phi$; instead, it is probably due to ringing.
 As in a Fourier decomposition, sharp changes in height will cause ringing in all frequencies of a wavelet transform. Indeed, the height map (Fig.~\ref{fig:height}) does have an abrupt drop near $\phi\approx90\degr$ which could cause this ringing; we mark this region with an `X'. The same feature can be seen after the warp has been subtracted in the ring at 28 kpc plotted in Figure \ref{fig:specring}. 
 
We use the local and global analyses in conjunction to  determine whether there is any scalloping mode that is present over a full $2\pi$ ring in the outer Galaxy. This is important because it will determine whether the mechanism that causes scalloping operates on a global or local scale. For this task, both techniques are necessary because even a pure $m=10$ oscillation will have some width in frequency space when put through a wavelet transform. However, this same oscillation will have a sharp peak  in the Lomb periodogram.  Therefore, in order to state that some mode exists over a full $2\pi$, we require both significant strength in the Lomb periodogram and significant power over a large range of $\phi$ in the wavelet transform. This combination will conclusively determine whether the scalloping is a global or local phenomenon. 

For modes with $m\le6$ this technique is not useful. These modes are clearly important because their Lomb periodograms  show  significant power for $3\le m\le 6$. Unfortunately, these modes have large enough wavelengths that the excluded regions always interfere with the wavelet transform regardless of where in the disk we look. A method of reducing the size of the excluded regions would alleviate this problem. However, this is likely to be difficult because eliminating the excluded regions would require a detailed knowledge of both $v_R$ and the distribution of gas velocities due to turbulence in the disk. Portions of these region are  optically thick, which will make density determination impossible.
 
 With these constraints, the Lomb periodogram leads us to conclude that the modes $m=10$ and 15 are the most fruitful places to look for scalloping that travels a full circle around the disk. Other modes do have interesting features, but their smaller Lomb amplitudes imply that we could be fooled by the imperfect frequency resolution of the wavelet power spectrum. For example, the modes $7\le m\le9$ have significant wavelet power for large portions of the northern half of the galaxy, but none of these modes has significant signal for the corresponding radius range in the Lomb periodogram.

For the $m=10$ mode the only large region with significant power in the south is the one around $\ell\approx310\degr$ that we noted previously. The power in this perturbation has fallen below the 95\% confidence level by $\ell\approx270\degr$, implying that this scalloping is a local effect. 

The same appears to be true for the $m=15$ oscillations, although this mode does have a region of high significance near $R\approx30$ kpc over a large range of $\phi$. However, the northern part of this signal is the region where we believe ringing to play an important role. It is therefore possible that the $m=15$ mode does have significant wavelet power over the full $2\pi$ near $R\approx 30$ kpc, but the evidence is not conclusive because the northern part of the wavelet power is most likely not due to scalloping. 
 
 With the exception of the $m=15$ mode, no frequency we examine with the wavelet transform carries significant power around an entire ring along with a correspondingly significant Lomb periodogram strength. For this reason, we conclude that  scalloping generally appears to be a local phenomenon. The modes $3\le m \le 6$ remain a possible exception, since we were not able to study them with a wavelet transform.

 It remains unclear what mechanism acts as the energy source for the scalloping. One possible cause is a massive object passing through the disk that excites local vertical oscillations in the HI gas.  Azimuthally traveling wavefronts can be created by the  magnetic field threaded through the disk; if the field is primarily azimuthal in nature, vertical oscillations will have larger phase and group velocities in the azimuthal direction than in the radial direction. This could lead to scalloping such as that seen towards $\ell\approx 310\degr$.
  
\section{Conclusions}

We fit the global shape of the warp on our grid of concentric rings using a vertical offset and two sinusoidal modes. Outside of $R\approx20$ kpc, each of these three modes has  more power than any of the higher frequencies we look at. The amplitude increases with radius over our entire radius range for the 1 mode, and starting from around 15 kpc for the 0 and 2 modes. The growth of the 0 and 2 modes results in  asymmetry in the warp; this growth begins near where the stellar disk ends.
The line of maxima of the $m=1$ mode is essentially coincident with one of the lines of maxima of the $m=2$ mode. There is little evidence for precession or winding of these two modes.

A global analysis with the Lomb periodogram shows that each $m$ mode evolves differently with radius. The most interesting include $m=3-6,10$, and 15, each of which start at small radii below the 95\% significance level and then cross it further out in the disk.  An analysis combining the global Lomb periodogram and the local wavelet transform shows that none of the modes with $7\le m\le15$  have strength over a full ring of the disk. Using a wavelet transform, we show that the scalloping observed by previous authors near $\ell\sim 310\degr$ is real. We therefore conclude that the scalloping is a local effect.
 Lower frequency modes proved impossible to study with wavelets due to the presence of the excluded regions.

\acknowledgments

 Wavelet software was provided by C.~Torrence and G.~Compo and is available at URL: http://paos.colorado.edu/research/wavelets/ \citep{TC1998}. Many thanks to Peter Kalberla for providing a copy of the LAB data set.   Thanks to Eugene Chiang for discussions of dynamics. ESL and LB are supported by NSF grant AST 02-28963. CH is supported by NSF grant AST 04-06987.

\appendix
\section{Measuring radial motion of the gas}

\begin{figure*}
\includegraphics[angle=90,scale=.88]{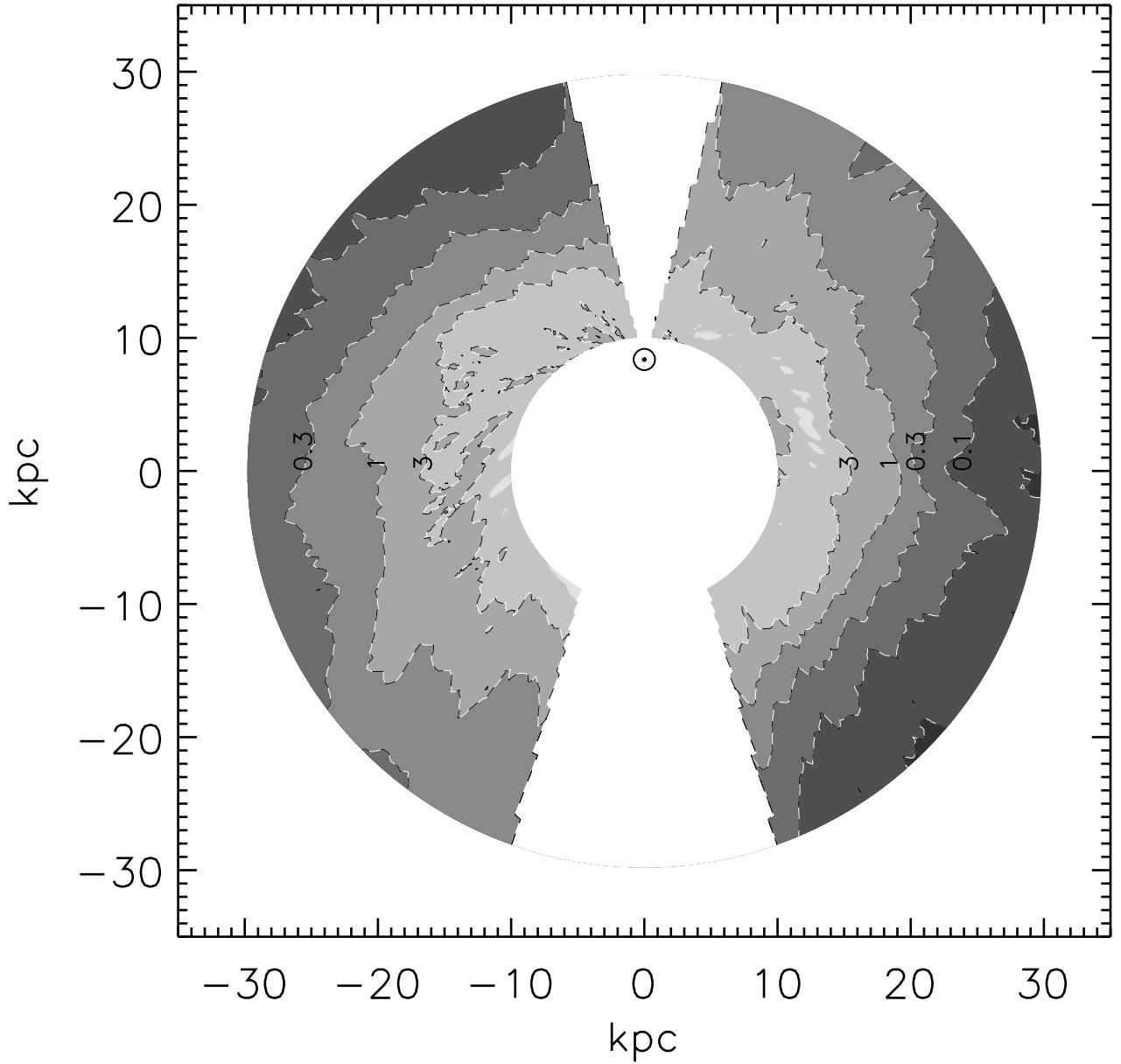}
\caption{\label{fig:novlsr} A contour map of $\Sigma(R,\phi)$ without any correction for elliptical orbits. Contours are evenly spaced in log$_{10}$ space, at $(3,1,0.3,0.1)~\mathrm{M}_\odot$/pc$^2$. }
\end{figure*}

This appendix describes our method for finding a functional form for $v_\Pi(R)$, the magnitude of the elliptical corrections to circular rotation. A naive surface density plot using only circular rotation has a large degree of asymmetry across the lines $\ell=0\degr$ and $180\degr$ \citep{KBH1982}. Figure \ref{fig:novlsr} shows the calculated surface density without any correction for elliptical gas orbits, but with the filters we describe in \S\ref{sec:proc}. Observational experience tells us that the surface density should not have discontinuities in cardinal directions; the elliptical orbits we describe in \S \ref{sec:proc} provide the strongest corrections in the directions where the $\cos \phi$ term in (\ref{eqn:transform}) is large, and smaller corrections elsewhere.  The algorithm is based on matching the surface density of the HI disk on either side of the excluded region centered at $\ell = 180\degr$. We draw confidence from the fact that this fit also does a good job matching the contours around $l=0\degr$ even though they are not included in the fit.

In order to calculate a surface density, we must first know $v_\Pi(R)$ because it enters into the derivative $|dv_r/dr|$. Thus, determining  $v_\Pi(R)$ from anything connected to $\rho$ is a circular problem.  We must assume some form of $v_\Pi(R)$ and check to see if it results in surface densities that are well matched. We choose the functional form:
 \begin{equation}
 v_\Pi(R)=\alpha\frac{(R-R_0)}{R_0}+\beta\frac{(R-R_0)^2}{R_0^2},
 \end{equation} 
 and do not include a zeroth order term to ensure that $v_\Pi(R)$ passes through zero for the solar circle. \citet{RS1980} showed that the LSR does not have a radial velocity with respect to the Galactic center.

Our algorithm follows these steps:
\begin{enumerate}
\item Construct $v_\Pi(R)$ for some combination of $\alpha$ and $\beta$
\item Interpolate from the LAB survey to find $T_b(R,\ell,z)$ for a grid in $R_i$ and $z_j$ over the survey range $\left|b\right| \le 30\degr$ and $155\degr \le \ell \le 165\degr$ or $195\degr\le \ell \le205\degr$ using $v_\Pi(R)$
\item Sum over $z$ to find the surface density $\Sigma(R,\ell)$
\item Average over $\ell$ for the two subsets of $\ell$ to find $\Sigma_{165}(R)$ and $\Sigma_{195}(R)$
\item Calculate a modified $\chi^2$ statistic to determine how well the surface densities are matched
\end{enumerate}

The modified $\chi^2$ is defined as
\begin{equation}
\chi^2=\sum_{R_i} \left[\frac{\Sigma_{165}(R_i)-\Sigma_{195}(R_i)}{\Sigma_{165}(R_i)<\Sigma_{195}(R_i)}\right]^2
\end{equation}
where the $<$ operator returns the smaller of its two operands.

Using this algorithm we search for the values of $\alpha$ and $\beta$ that minimize the modified $\chi^2$. We find good matches for $\alpha=8.67$ and $\beta=-1.08$.

We also tried a weighting that only  fit points beyond the perturbations of the spiral arms, i.e. $R>2R_0$. Although this did change $v_\Pi(R)$ by 25\% or so, it had no qualitative effect on our other results.

\begin{figure}
\epsscale{1.2}
\plotone{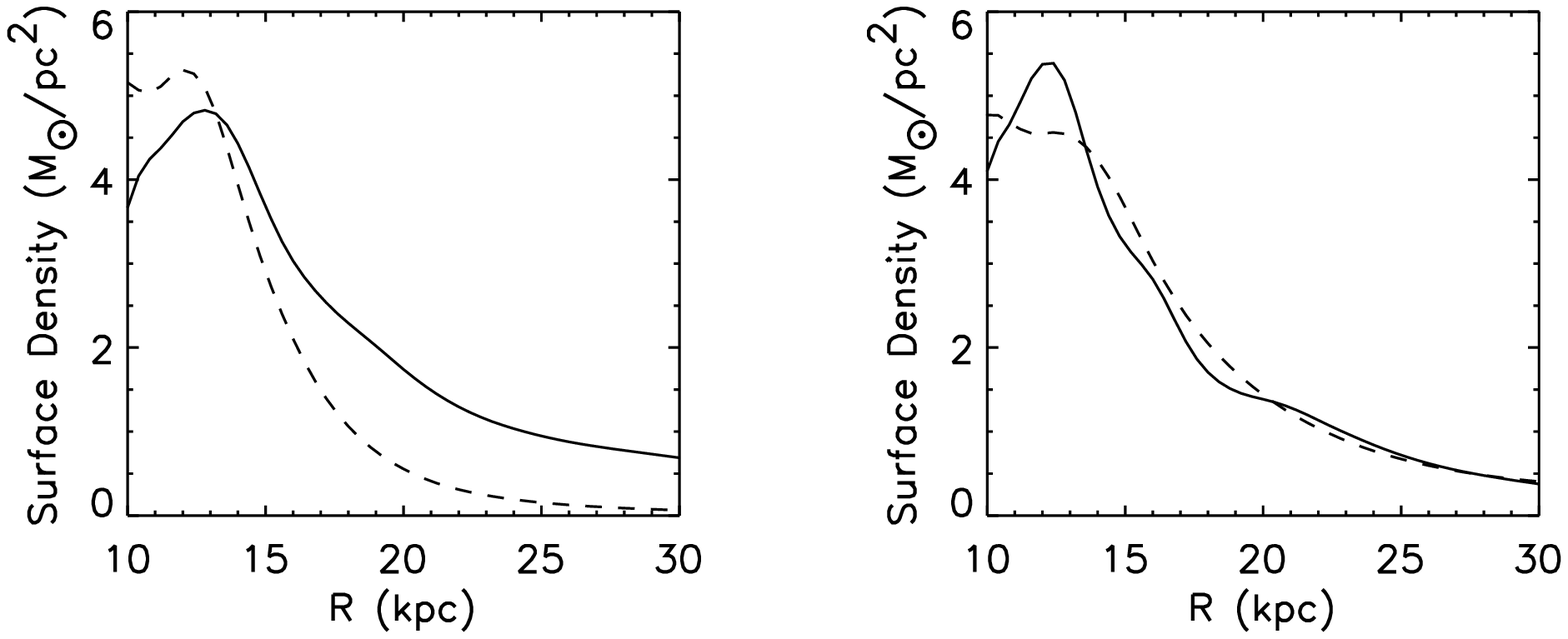}
\caption{\label{fig:sigmatch} Left: The average surface densities with no correction for $v_\Pi(R)$.  The solid line is $\Sigma_{165}(R)$ and the dotted line is $\Sigma_{195}(R)$. Right: The average surface densities after being adjusted with $v_\Pi(R)$.}
\epsscale{1}
\end{figure}

\section{Wavelets}

The continuous wavelet transform for a discrete series of points $z_n$ is given by
\begin{equation}\label{eqn:wavetrans}
W_n(s)=\sum^{N-1}_{n'=0} z_{n'} \psi^*\left[
\frac{(n'-n)\Delta\phi}{s}\right]
\end{equation}
where $N$ is the number of points in the series, $n$ is a position index,  $\Delta\phi$ is the spacing of the points in $\phi$ space, and $\psi^*$ is the complex conjugate of the normalized Morlet wavelet \citep{TC1998}:  
\begin{equation}
\psi\left(\eta\right)
=\left(\frac{\Delta\phi}{s}\right)^{1/2}\pi^{-1/4}e^{i\omega_0\eta}e^{-\eta^2/2}.
\end{equation}
The scale of the transform is $s$; for the Morlet wavelet this is simply related to the wavelength by $\lambda=1.03s$. Evenly spaced points are necessary in order to use this transform, but since we are treating the two halves of the Galaxy separately, $s(R,\phi)$ is  split into two halves, each with equally spaced points.  We perform this transform for the dense set of frequencies given by 
\begin{equation}
s_j=s_02^{j\Delta j}, ~~~j=0,1,\ldots,J.
\end{equation}
Here, $s_0$ is the smallest scale that can be sampled, $2\Delta\phi$. $\Delta j$ is a measure of how densely we sample in  scale space; because computation time is not large, we  choose relatively dense sampling throughout: $\Delta j = 0.0125$. 

From (\ref{eqn:wavetrans}), we construct the wavelet power spectrum, $\left|W_n(s)\right|^2$.
We calculate the wavelet power spectrum for the dense set of frequencies, and then filter over a range of scales to find the power in a frequency band. The filtered power spectrum is given by
\begin{equation}
\overline{W_n}^2=\frac{\Delta j\Delta \phi}{C_\delta}\sum_{j=j_1}^{j_2}\frac{\left|W_n(s_j)\right|^2}{s_j}
\end{equation}
where $C_\delta$ is a reconstruction factor that depends on the choice of wavelet. For the Morlet wavelet with $\omega_0 =6$, $C_\delta=0.776$.


\begin{thebibliography}{54}
\bibitem[{{Burton}(1988)}]{B1988}
---. {The structure of our Galaxy derived from observations of neutral
  hydrogen} (Galactic and Extragalactic Radio Astronomy), 295--358
\end{thebibliography}

\end{document}